\documentclass[11pt,a4paper]{article}

\usepackage[T1]{fontenc}
\usepackage{lmodern}
\usepackage{microtype}
\usepackage{geometry}
\usepackage{setspace}
\usepackage{amsmath}
\usepackage{graphicx}
\usepackage{booktabs, appendix}
\usepackage[round,authoryear]{natbib}
\usepackage{hyperref}   
\usepackage{adjustbox}
\usepackage{algorithm}
\usepackage{algpseudocode}
\usepackage{amsfonts, comment}
\usepackage[table]{xcolor}
\usepackage{float}
\usepackage{multirow}
\usepackage{makecell}

\begin{document}

\begin{titlepage}
    \centering

    \vspace*{3cm}

    {\textbf{\LARGE
    Score-optimized copula coupling for\\[0.2cm]
    multivariate ensemble post-processing
    }}

    \vspace{2cm}

    {\large
    Mária Lakatos\textsuperscript{*}
    \par}

    \vspace{0.8cm}

    {\normalsize
    Faculty of Informatics, University of Debrecen\\
    Kassai út 26, H-4028 Debrecen, Hungary
    \par}

    \vspace{0.5cm}

    {\small
    \textsuperscript{*}Corresponding author:
    \texttt{lakatos.maria@inf.unideb.hu}
    \par}

    \vspace{2.3cm}

\begin{abstract}
Statistical post-processing can calibrate the individual components of an ensemble forecast but may also distort relationships between forecast dimensions. This paper proposes Score-Optimized Copula Coupling, which rearranges calibrated forecast samples according to an optimized dependence structure to restore potentially lost dependencies. We assess fixed and adaptively updated versions using wind-speed and temperature forecasts. In the variable-specific spatial case studies, all versions significantly outperform the strongest benchmarks, with improvements of 1.08-2.02 percent in energy score and 5.45-7.97 percent in variogram score. A third case study at 24 co-located stations shows that one optimized structure can represent both spatial and inter-variable relationships. The largest energy and variogram score improvements are 3.98 and 3.69 percent for spatial dependence and 1.16 and 1.06 percent for inter-variable dependence; all are statistically significant. Additional calibration terms improve covariance representation but can reduce gains in the other scores.
\end{abstract}


\end{titlepage}

\section{Introduction}

Calibrated marginal predictive distributions and an appropriate representation of the dependence structure across dimensions are key characteristics of reliable multivariate probabilistic forecasts. Accurate modeling of these dependencies is essential in numerous applications in which decisions depend on the joint occurrence of events rather than on their marginal behavior alone. For instance, forecasts of power generation and demand may each be well calibrated, but an incorrect representation of their dependence can substantially underestimate the probability of supply-demand imbalances \citep{stover2024dependence}. In aviation meteorology, spatially coherent predictions are required to represent adverse visibility conditions across multiple locations \citep{lakatos2024visibility}.

The introduction of ensemble prediction systems (EPSs) into operational weather forecasting marked an important milestone in meteorology. However, both physics- and artificial-intelligence-based weather prediction systems are subject to systematic biases, and their forecasts are therefore often uncalibrated \citep{GneitingEtAl2026}. Since the early 2000s, numerous statistical and machine-learning-based post-processing methods have been proposed to address these deficiencies, including ensemble model output statistics \citep[EMOS;][]{GneitingEtAl2005}, Bayesian model averaging \citep[BMA;][]{RafteryEtAl2005}, and, more recently, distributional regression networks \citep[DRN;][]{RaspLerch2018}.

Separate marginal calibration may, however, distort or remove the dependence structure across forecast dimensions. Empirical-copula-based multivariate post-processing methods are among the most widely studied approaches for restoring this dependence. Samples or quantiles from the calibrated marginal predictive distributions are rearranged according to a dependence template, which may be obtained from the rank structure of raw ensemble forecasts, historical observations, or forecast-error correlations \citep[e.g.,][]{ClarkEtAl2004,SchefzikEtAl2013,BenBouallegueEtAl2016}. Although these methods are generally computationally efficient, studies covering different weather variables and prediction systems have not identified a uniformly superior approach. Moreover, the quality of the dependence information contained in the raw ensemble may strongly affect methods that use its rank structure as a dependence template \citep{LakatosEtAl2023,Lakatos2026}.

An alternative is to model the joint predictive distribution directly \citep[e.g.,][]{SchuhenEtAl2012,SloughterEtAl2013,BaranMoller2017}. While these models can simultaneously calibrate the marginal distributions and represent their dependence structure, a limitation is the need to identify a suitable parametric distribution family for variables with different characteristics or for forecasts over large station networks.

Machine learning provides another route to multivariate post-processing. Instead of prescribing the dependence structure through a fixed parametric family, these models can learn nonlinear relationships from large sets of predictors. Generative models, graph neural networks, and multivariate distributional regression networks have been considered for this purpose \citep[e.g.,][]{DaiHemri2021,ChenEtAl2024,Lakatos2026,CasellasEtAl2026}. Their flexibility, however, usually comes with greater data and computational requirements, as well as additional choices concerning model architecture and hyperparameter tuning.

\begin{figure}[t]
    \centering
    \includegraphics[width=0.9\linewidth]{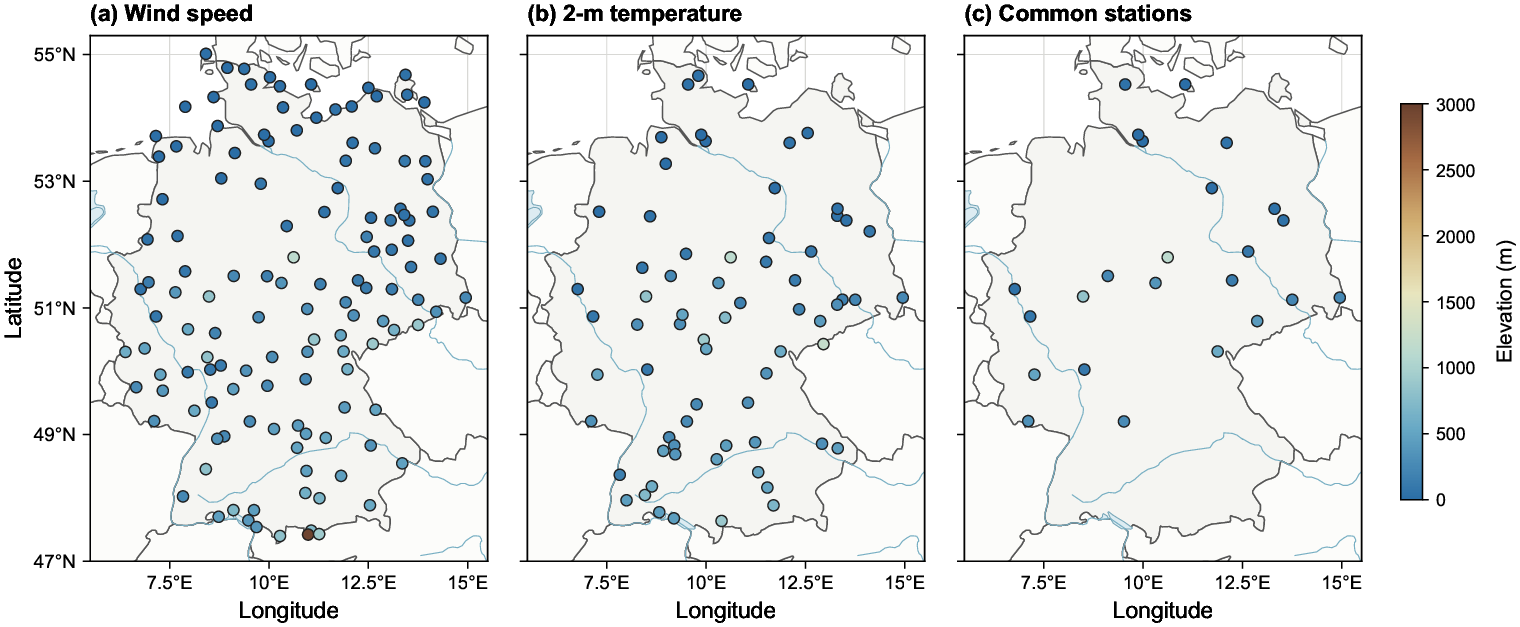}
    \caption{Locations of the stations used in the (a) wind speed, (b) 2-m temperature, and (c) joint spatial and inter-variable case studies.}
    \label{fig:map}
\end{figure}

This paper proposes Score-Optimized Copula Coupling (SOCC), a general nonparametric multivariate post-processing method that follows the two-stage framework of empirical-copula-based approaches. In the first stage, calibrated univariate forecasts are obtained. In the second stage, samples from these forecasts are reordered according to a rank template estimated by minimizing a loss function designed to improve the representation of the multivariate dependence structure.

Starting from an initial dependence template constructed from raw ensemble forecasts or equidistant quantiles of calibrated marginal predictive distributions, SOCC optimizes the rank structure through a sequence of rank rearrangements. The resulting score-optimized template is then used to reorder equidistant quantiles of the post-processed marginal distributions during the validation period. We consider both static and warm-start variants. The static approach uses a single optimized template for all subsequent forecast cases, whereas the warm-start approach updates the template using forecast--observation pairs from a rolling training window preceding each validation date.

SOCC is evaluated using 2-m temperature and wind-speed ensemble forecasts. We first investigate its ability to restore spatial dependencies that may have been distorted by separate marginal calibration. We then assess whether SOCC can simultaneously represent spatial and inter-variable dependencies between temperature and wind speed. The benchmark methods include the Gaussian copula approach \citep{MollerEtAl2013}, the more recent COBASE method proposed by \citet{FlosEtAl2026}, ensemble copula coupling and its variants \citep{SchefzikEtAl2013,BenBouallegueEtAl2016}, and several variants of the Schaake shuffle \citep{ClarkEtAl2004,Schefzik2016,ScheuererEtAl2017}.

Although SOCC is evaluated using wind-speed and temperature forecasts, its formulation is not specific to meteorology or to a particular marginal calibration method. In principle, it can be applied to continuous or discrete predictive distributions, with forecast dimensions representing different locations, variables, lead times, or combinations thereof. Furthermore, the initial rank template may be constructed from any available source of dependence information and is not restricted to the raw-ensemble and historical-observation templates commonly used by ECC and the Schaake shuffle.

The remainder of the paper is organized as follows. Section \ref{sec2} describes the wind-speed and temperature data used in the case studies. Section \ref{sec3} introduces the proper scoring rules and multivariate rank diagnostics used to evaluate the forecasts. Section \ref{sec4} presents the applied EMOS models (Section \ref{sec4.1}), the proposed SOCC method (Section \ref{sec4.2}), its implementation details (Section \ref{sec4.3}), and the benchmark multivariate post-processing approaches (Section \ref{sec4.4}). Section \ref{sec5} reports the results of the three case studies, followed by a discussion and conclusions in Section \ref{sec6}.

\section{Data}
\label{sec2}

The study uses the dataset published by \citet{ChenEtAl2024}, which contains 50-member ensemble forecasts from the European Centre for Medium-Range Weather Forecasts (ECMWF) for 2-m temperature and 10-m wind speed over Germany. The forecasts are initialized once daily at 00 UTC and have a lead time of 48 hours. They were obtained from the THORPEX Interactive Grand Global Ensemble (TIGGE) archive \citep{BougeaultEtAl2010}, in which ECMWF forecasts are available on a \(0.5^{\circ}\times0.5^{\circ}\) grid. The gridded forecasts were interpolated to the locations of the meteorological stations. Corresponding observations were provided by the Climate Data Center of the German Meteorological Service (Deutscher Wetterdienst, DWD).

The complete archive covers the period from 3 January 2007 to 31 December 2016. The present analysis uses 129 wind-speed stations and 69 temperature stations for which both forecasts and observations are available. The temperature observations are complete, whereas less than \(1\%\) of the wind-speed observations are missing. Forecast cases with incomplete data for any station were excluded. The experiments focus on 2016, with forecast--observation pairs from 2 April to 31 December 2016 used as the validation period. Earlier years of the archive were retained as historical data for the multivariate post-processing methods described in Sections \ref{sec4.2} and \ref{sec4.4}. Figure \ref{fig:map} shows the wind-speed and temperature station networks together with the co-located stations used in the joint case study.

\section{Verification}
\label{sec3}

In the case studies of Section \ref{sec5}, the performance of the univariate forecasts was evaluated using the continuous ranked probability score \cite[CRPS;][]{Wilks2019}, a proper scoring rule that assesses calibration and sharpness simultaneously. For a predictive cumulative distribution function \(F\) and observation \(y\), the continuous CRPS is
\[
\operatorname{CRPS}(F,y)
=
\mathbb{E}|X-y|
-\frac{1}{2}\mathbb{E}|X-X'|,
\]
where \(X\) and \(X'\) are independent random variables with distribution \(F\). For \(\boldsymbol{f}=(f_1,\ldots,f_K)\), a \(K\)-member ensemble forecast, the sample CRPS is:
\[
\operatorname{CRPS}(\boldsymbol{f},y)
=
\frac{1}{K}\sum_{k=1}^{K}\left|f_k-y\right|
-
\frac{1}{2K^2}
\sum_{k=1}^{K}\sum_{\ell=1}^{K}
\left|f_k-f_\ell\right|.
\]
The CRPS is negatively oriented, meaning that lower CRPS values indicate better probabilistic forecasts. To assess the skill of multivariate predictions, we used the energy score \citep[ES;][]{GneitingRaftery2007} and the variogram score of order \(p\) \citep[VS;][]{GneitingEtAl2008}. Let \(\boldsymbol{f}_1,\ldots,\boldsymbol{f}_K\in\mathbb{R}^{D}\) denote the \(K\) ensemble members, let \(F_K\) denote their empirical distribution, and let \(\boldsymbol{y}\in\mathbb{R}^{D}\) be the verifying observation. The energy score is
\[
\operatorname{ES}(F_K,\boldsymbol{y})
=
\frac{1}{K}\sum_{k=1}^{K}
\left\|\boldsymbol{f}_k-\boldsymbol{y}\right\|
-
\frac{1}{2K^2}
\sum_{k=1}^{K}\sum_{\ell=1}^{K}
\left\|\boldsymbol{f}_k-\boldsymbol{f}_{\ell}\right\|,
\]
where \(\|\cdot\|\) denotes the Euclidean norm. The variogram score of order \(p\) is
\[
\operatorname{VS}^{p}(F_K,\boldsymbol{y})
=
\sum_{i=1}^{D}\sum_{j=1}^{D}
w_{ij}
\left(
\left|y^{(i)}-y^{(j)}\right|^{p}
-
\frac{1}{K}\sum_{k=1}^{K}
\left|f_{k}^{(i)}-f_{k}^{(j)}\right|^{p}
\right)^{2},
\]
where \(w_{ij}\geq 0\) denotes the weight assigned to the coordinate pair \((i,j)\). While the ES, the multivariate generalization of the CRPS, quantifies forecast quality in terms of deviation from the observation vector, the VS is more sensitive to misspecification of the correlation structure. In this paper, we use \(p=1\) and equal weights for all station pairs, both during forecast evaluation and when training the SOCC algorithm discussed in Section \ref{sec4.2}. The CRPS, ES, and VS were computed using the corresponding implementations in the \texttt{R} package \texttt{scoringRules} \citep{JordanEtAl2019}. Moreover, we considered the Dawid--Sebastiani score (DSS), which is based on the first two moments of the predictive distribution. For a \(D\)-dimensional ensemble forecast \(\boldsymbol{f}_1,\ldots,\boldsymbol{f}_K\) and observation vector \(\boldsymbol{y}\), the DSS is defined as
\[
\operatorname{DSS}(F_K,\boldsymbol{y})
=
\log\det(\boldsymbol{S})
+
(\boldsymbol{y}-\overline{\boldsymbol{f}})^\top
\boldsymbol{S}^{-1}
(\boldsymbol{y}-\overline{\boldsymbol{f}}),
\]
where \(\overline{\boldsymbol{f}}\) and \(\boldsymbol{S}\) denote the ensemble mean vector and sample covariance matrix, respectively. To handle the possible singularity of \(\boldsymbol{S}\), we use the regularized DSS considered by \citet{Wilks2020}:
\[
\operatorname{RDSS}(F_K,\boldsymbol{y})
=
\log\det\!\left\{\boldsymbol{\Theta}^{-1}(\lambda)\right\}
+
(\boldsymbol{y}-\overline{\boldsymbol{f}})^\top
\boldsymbol{\Theta}(\lambda)
(\boldsymbol{y}-\overline{\boldsymbol{f}}),
\]
where \(\boldsymbol{\Theta}(\lambda)\) is a regularized estimate of the precision matrix obtained using the graphical lasso, and \(\lambda\) controls the strength of the regularization and hence the sparsity of the estimated precision matrix. We use the prespecified value \(\lambda=0.02\) for all forecast cases and methods in all three case studies. Thus, differences in the RDSS cannot be attributed to forecast-specific or method-specific selection of the regularization parameter.

For a forecast \(F\), one can also quantify the improvement in a score \(S\) with respect to a reference forecast \(F_{\mathrm{ref}}\). These skill scores are defined as
\[
\operatorname{SS}
=
100
\left(
1-\frac{\overline{S}(F)}
        {\overline{S}(F_{\mathrm{ref}})}
\right)\%,
\]
where \(\overline{S}\) denotes the mean score over the validation period. In the case studies of Section \ref{sec5}, we consider the energy skill score (ESS), variogram skill score (VSS) and RDSS skill score, obtained by using the ES, VS and RDSS, respectively. Positive skill scores indicate an improvement over the reference forecast.

Multivariate calibration was additionally assessed using the average-rank, band-depth, energy-rank, and dependence-rank histograms proposed and discussed by \citet{AllenEtAl2024}. The corresponding pre-rank functions are implemented in the accompanying \texttt{R} package \texttt{MultivCalibration}. Each histogram targets a different characteristic of the multivariate forecast distribution, therefore, a nearly uniform histogram for one rank type does not necessarily imply overall multivariate calibration. Deviations from uniformity were quantified using the reliability index \citep[RI;][]{DelleMonacheEtAl2006}:
\[
\operatorname{RI}
=
\sum_{r=1}^{K+1}
\left|
\rho_r-\frac{1}{K+1}
\right|,
\]
where \(\rho_r\) denotes the relative frequency of rank \(r\). Smaller RI values indicate a rank distribution closer to the discrete uniform distribution.

Finally, to investigate the statistical significance of score differences between two competing forecasts \(F\) and \(G\), Diebold-Mariano test statistics \citep[DM test;][]{DieboldMariano1995} were calculated, following \citet{GneitingRanjan2011} and \citet{LakatosEtAl2023}. Under the null hypothesis of equal predictive performance, the DM test statistic \(t_N\) asymptotically follows the standard normal distribution, which is defined as:
\[
t_N
=
\sqrt{N}\,
\frac{\overline{S}(F)-\overline{S}(G)}
     {\widehat{\sigma}_N},
\]
where \(\overline{S}(F)\) and \(\overline{S}(G)\) are the mean scores over the validation period for forecasts \(F\) and \(G\), respectively, and \(\widehat{\sigma}_N\) is an estimator of the asymptotic standard deviation of the sequence of individual score differences. Negative values of \(t_N\) favour \(F\), whereas positive values favour \(G\) \citep{LakatosEtAl2023}.

\section{Post-processing methods}
\label{sec4}

This section describes the univariate and multivariate post-processing methods applied to the temperature and wind speed forecasts. First, we present the ensemble model output statistics approach used for marginal calibration. We then introduce the Score-Optimized Copula Coupling algorithm and its implementation details, followed by the benchmark multivariate post-processing methods.

\subsection{Ensemble model output statistics}
\label{sec4.1}

Ensemble model output statistics \citep[EMOS;][]{GneitingEtAl2005} is a widely used statistical approach for calibrating ensemble forecasts. It corrects biases and dispersion errors in the raw ensemble by minimizing the CRPS, as defined in Section \ref{sec3}. The parameters of the post-processed predictive cumulative distribution function (CDF) are expressed as affine functions of summary statistics of the raw ensemble, with the corresponding coefficients estimated from historical forecast-observation pairs. For both wind speed and 2-m temperature, we consider the following location-scale model:
\[
Y\mid\mathbf{f}\sim\mathcal{D}(\mu,\sigma^2),
\qquad
\mu=a_0+a_1\bar{f},
\qquad
\sigma^2=b_0+b_1S^2,
\]
where \(Y\) denotes the future observation,
\(\mathbf{f}=(f_1,\ldots,f_K)^\top\) is the raw ensemble forecast, and
\[
\bar{f}=\frac{1}{K}\sum_{k=1}^{K}f_k,
\qquad
S^2=\frac{1}{K-1}\sum_{k=1}^{K}(f_k-\bar{f})^2
\]
are the ensemble mean and ensemble variance, respectively, while $a_0,a_1,b_0$, and $b_1$ are the estimated coefficients. For 2-m temperature, we employ a Gaussian predictive distribution, which
is commonly used in statistical post-processing of temperature forecasts
\citep{GneitingEtAl2005,RafteryEtAl2005}. Since wind speed is non-negative
and typically right-skewed, its predictive distribution is represented by
a normal distribution left-truncated at zero,
$\mathcal{N}_{0}^{\infty}(\mu,\sigma^2)$, following the suggestion of
\citet{thorarinsdottir2010probabilistic}. In the case studies of Section \ref{sec5}, regional estimation was applied, meaning that a single set of EMOS parameters was estimated jointly for the entire ensemble domain, using a rolling training period of 35 days for wind speed and 30 days for 2-m temperature.

\subsection{Score-optimized copula coupling}
\label{sec4.2}

Let \(\mathcal{T}_{\mathrm{train}}\) denote the historical training
period, consisting of past forecast-observation pairs. For each training day \(t\in\mathcal{T}_{\mathrm{train}}\) and
station \(d\), we consider \(K\) ordered forecast values
\[
f_{t,d,(1)}\leq\cdots\leq f_{t,d,(K)}.
\]
Here, we consider two variants of SOCC depending on the selection of marginal forecast values, subject to reordering. In the first case, the rank permutation is based on the raw ensemble forecasts (referred to as SOCC-RAW), while SOCC-EMOS and SOCC-Combined utilize equidistant quantiles of calibrated predictive CDFs, which are specifically retained for training.

The goal of SOCC is to estimate a rank template \(
\mathbf{R}\in\{1,\ldots,K\}^{D\times K},
\) which is based on the minimization of weighted scoring rules and/or multivariate rank penalties. Once the template is estimated, all future predictions in the validation set, that are equidistant quantiles of the predictive EMOS CDFs are rearranged to the rank order of \(\mathbf{R}\).
In the case of the SOCC-RAW and SOCC-EMOS variants, the loss function is the combination of the ES and VS (discussed in Section \ref{sec3}), defined as:
\[
\mathcal{L}_{\mathrm{SOCC}}(\mathbf{R})
=
\alpha
\frac{\overline{\mathrm{ES}}(\mathbf{R})}
     {\overline{\mathrm{ES}}(\mathbf{R}_{\mathrm{init}})}
+
(1-\alpha)
\frac{\overline{\mathrm{VS}}(\mathbf{R})}
     {\overline{\mathrm{VS}}(\mathbf{R}_{\mathrm{init}})},
\]
where \(\alpha\in[0,1]\) quantifies the relative contribution of each multivariate score, \(\overline{\mathrm{ES}}(\mathbf{R})\) and \(\overline{\mathrm{VS}}(\mathbf{R})\) are the mean ES and VS values, obtained after applying the rank structure \(\mathbf{R}\) to the training forecasts, while \(\overline{\mathrm{ES}}(\mathbf{R}_{\mathrm{init}})\) and \(\overline{\mathrm{VS}}(\mathbf{R}_{\mathrm{init}})\) are the mean ES and VS values obtained based on the initial rank template for the same training data. Hence, ratios below one for each term represent improvement relative to the initial state.

As the number of \((K!)^D\) templates that could be found by such a permutation of ranks is extensive (especially in higher dimensions), the SOCC algorithm employs a local search procedure described in Algorithm \ref{alg:socc}. As a result, the optimization step is performed only once (therefore estimating a static template), substantially decreasing computational time. Alternatively, one could also use a rolling training period, however, as a compromise between static and rolling training, we utilize warm-start variants instead. In this case, building on the optimized static \(\mathbf{R}\) of SOCC, a slight update of ranks is permitted based on past forecast-observation pairs closer to the actual validation dates, therefore providing the possibility of introducing time-dependent information to \(\mathbf{R}\).

\begin{algorithm}[t]
\small
\caption{Score-Optimized Copula Coupling (SOCC)}
\label{alg:socc}
\begin{algorithmic}[1]
\Require Sorted marginal training forecasts
\(\{f_{t,d,(k)}\}\), observations \(\{y_{t,d}\}\), initial rank
template \(\mathbf{R}_{\mathrm{init}}\), number of proposals \(P\),
and loss function \(\mathcal{L}\)

\State Set \(\mathbf{R}\gets\mathbf{R}_{\mathrm{init}}\)
\State Evaluate \(L\gets\mathcal{L}(\mathbf{R})\)

\For{\(p=1,\ldots,P\)}
    \State Select a station \(d\) uniformly at random
    \State Select distinct positions \(k_1,k_2\in\{1,\ldots,K\}\)
    \State Construct \(\mathbf{R}'\) by exchanging
    \(R_{d,k_1}\) and \(R_{d,k_2}\)
    \State Rearrange the training forecasts according to \(\mathbf{R}'\)
    \State Evaluate \(L'\gets\mathcal{L}(\mathbf{R}')\)
    \If{\(L'<L\)}
        \State Set \(\mathbf{R}\gets\mathbf{R}'\) and \(L\gets L'\)
    \EndIf
\EndFor

\State Set \(\mathbf{R}^{*}\gets\mathbf{R}\)

\If{warm-start rolling updates are used}
    \For{each forecast date \(s\)}
        \State Use \(\mathbf{R}^{*}_{s-1}\) as the initial template
        \State Update the template using the most recent \(W\) training
cases available before forecast date \(s\) and \(P_{\mathrm{u}}\) proposals
        \State Store the resulting template as \(\mathbf{R}^{*}_{s}\)
    \EndFor
    \State \Return \(\{\mathbf{R}^{*}_{s}\}\)
\Else
    \State \Return \(\mathbf{R}^{*}\)
\EndIf
\end{algorithmic}
\end{algorithm}

However, when evaluating multivariate forecast quality, only considering the mean ES and VS scores is not necessarily enough, as ensembles with different rank-dependence structures can produce similar average scores. This motivates the use of an extended objective that, in addition to the energy and
variogram scores, penalizes departures from uniformity in the average-rank
and band-depth rank distributions:
\[
\mathcal{L}_{\mathrm{Combined}}(\mathbf{R})
=
(1-\beta_A-\beta_B)\mathcal{L}_{\mathrm{SOCC}}(\mathbf{R})
+
\beta_A
\frac{\mathcal{P}_{A}(\mathbf{R})}
     {\mathcal{P}_{A}(\mathbf{R}_{\mathrm{init}})}
+
\beta_B
\frac{\mathcal{P}_{B}(\mathbf{R})}
     {\mathcal{P}_{B}(\mathbf{R}_{\mathrm{init}})}.
\]
Here, \(\beta_A,\beta_B>0\) and \(\beta_A+\beta_B<1\) are the weights of the average rank \(\mathcal{P}_{A}(\cdot)\) and band-depth rank \(\mathcal{P}_{B}(\cdot)\) penalties, respectively. For a
given rank template, one rank \(r_{j,t}\) of each type is calculated for each
training day \(t\) and normalized as
\[
u_{j,t}
=
\frac{r_{j,t}-\tfrac{1}{2}}{K+1},
\qquad j\in\{A,B\},
\]
where \(K\) is the ensemble size. For \(T = \lvert\mathcal{T}_{\mathrm{train}}\rvert\) training days, \(\mathcal{P}_{j}\) is 
defined as
\[
\mathcal{P}_{j}
=
\frac{1}{T}\sum_{t=1}^{T}
\left(
u_{j,(t)}-\frac{t-\tfrac{1}{2}}{T}
\right)^2,
\qquad j\in\{A,B\},
\]
where \(u_{j,(t)}\) denotes the \(t\)th ordered normalized rank. Thus, the
penalty measures the squared distance of the empirical rank distribution from
uniformity, with smaller values indicating better rank calibration. A
normalized penalty below one indicates an improvement over the initial
template for the corresponding diagnostic.

\subsection{Implementation details}
\label{sec4.3}

\begin{table}[t]
\centering
\small
\setlength{\tabcolsep}{4pt}
\renewcommand{\arraystretch}{1.05}

\begin{tabular}{@{}llrrrrrrr@{}}
\toprule
Case study
& SOCC variant
& \(\lvert\mathcal{T}_{\mathrm{train}}\rvert\)
& \(\alpha\)
& \(\beta_A\)
& \(\beta_B\)
& \(W\)
& \(P_u\)
& \(\alpha_u\) \\
\midrule

\multirow{3}{*}{Wind speed (spatial)}
& SOCC-EMOS     & 30  & 0.50 & 0     & 0     & 60 & 100 & 0.50 \\
& SOCC-RAW      & All & 0.25 & 0     & 0     & 60 & 100 & 0.50 \\
& SOCC-Combined & 60  & 0.75 & 0.01  & 0.01  & 30 & 50  & 0.75 \\

\addlinespace[3pt]

\multirow{3}{*}{Temperature (spatial)}
& SOCC-EMOS     & 50  & 0.75 & 0     & 0     & 60 & 100 & 0.50 \\
& SOCC-RAW      & All & 0.25 & 0     & 0     & 60 & 100 & 0.50 \\
& SOCC-Combined & 50  & 0.75 & 0.005 & 0.005 & 60 & 100 & 0.50 \\

\addlinespace[3pt]

\multirow{3}{*}{%
  \makecell[l]{Temperature and wind speed\\
  (spatial and inter-variable)}}
& SOCC-EMOS     & 30  & 0.50 & 0    & 0    & 60 & 100 & 0.50 \\
& SOCC-RAW      & All & 0.75 & 0    & 0    & 60 & 100 & 0.50 \\
& SOCC-Combined & 30  & 0.75 & 0.01 & 0.01 & 60 & 100 & 0.50 \\

\bottomrule
\end{tabular}

\caption{Selected SOCC hyperparameters.
The static training period is denoted by
\(\lvert\mathcal{T}_{\mathrm{train}}\rvert\), with ``All'' indicating the
use of all available historical forecast--observation pairs up to the end
of the training period. The parameters \(\alpha\), \(\beta_A\), and
\(\beta_B\) denote the static ES weight and the average-rank and band-depth
penalty weights, respectively. The warm-start window length, number of
update proposals, and ES weight are denoted by \(W\), \(P_u\), and
\(\alpha_u\), respectively.}
\label{tab:socc_hyperparameters}
\end{table}

For all three case studies in Section \ref{sec5}, the SOCC templates based on rank optimization of equidistant EMOS quantiles were estimated using forecast-observation pairs from 1 January to 31 March 2016. SOCC-RAW constructs the dependence template by reordering raw ensemble forecasts and uses all available historical data. In all cases, the initial template \(\mathbf{R}_{\mathrm{init}}\) was obtained from the multivariate ranks of 50 historical observations, analogously to the dependence-template construction in the Schaake shuffle (SSh). Improvements in multivariate forecast quality were measured relative to this initial template. Based on the notation introduced in Algorithm \ref{alg:socc}, we set \(P=50{,}000\) for static template estimation. For the warm-start variants, the template was updated daily using only forecast-observation pairs available before the corresponding forecast date.

To make the contributions of the different loss components comparable, each component was normalized by its value under the initial rank template. As a result, the weights controlled their relative contributions independently of the original scales of the ES, VS, and multivariate rank penalties. In particular, setting \(\alpha=0.5\) assigned equal weight to the normalized ES and VS components. A brief discussion of the computational complexity of SOCC compared with GCA and ECC is provided in the Appendix, while Table \ref{tab:socc_computational_times} summarizes the runtimes of the SOCC variants applied in the case studies of Section \ref{sec5}.

We additionally investigate whether SOCC can simultaneously restore spatial
and inter-variable dependencies. For this purpose, we consider \(S=24\)
co-located stations for which both wind-speed and temperature
forecast--observation pairs are available (Figure \ref{fig:map}c). The two
variables are concatenated into a single forecast vector, giving \(D=2S=48\)
dimensions and a \(D\times K=48\times50\) SOCC rank template. Since wind speed
and temperature are expressed on different physical scales, each dimension
is standardized using its mean and standard deviation estimated from the
training observations. Both SOCC optimization and forecast verification are
performed using these standardized values.

The variogram-score component of the joint loss contains
\(\binom{S}{2}\) station pairs for each weather variable and \(S\) co-located
wind-speed--temperature pairs. Let
\[
G_{ij}
=
\left(
\left|z^{(i)}-z^{(j)}\right|^p
-
\frac{1}{K}\sum_{k=1}^{K}
\left|f_k^{(i)}-f_k^{(j)}\right|^p
\right)^2,
\]
where \(\boldsymbol{z}\) and \(\boldsymbol{f}_k\) denote the standardized
observation and the \(k\)th standardized ensemble member, respectively.
Indexing the wind-speed dimensions by \(1,\ldots,S\) and the corresponding
temperature dimensions by \(S+1,\ldots,2S\), the joint variogram score is
\[
\operatorname{VS}^{p}_{\mathrm{joint}}
=
2\left[
\sum_{1\leq s<t\leq S}G_{s,t}
+
\sum_{1\leq s<t\leq S}G_{S+s,S+t}
+
(S-1)\sum_{s=1}^{S}G_{s,S+s}
\right],
\qquad p=1.
\]
The first two terms assess the spatial dependence of wind speed and
temperature, respectively, whereas the third assesses their dependence at
the co-located stations. Each inter-variable pair receives a weight of
\(S-1\), since \(S(S-1)=2\binom{S}{2},\) which garantees equal weights of the components within the \(\operatorname{VS}^{p}_{\mathrm{joint}}\) calculation.  The ES is evaluated jointly over all \(D\) dimensions. The variable-specific training-window lengths and loss-function weights used for the SOCC variants are reported in Table \ref{tab:socc_hyperparameters}.

\subsection{Benchmark multivariate post-processing methods}
\label{sec4.4}

\begin{table}[t]
\centering
\begin{tabular}{lrrrr}
\toprule
Score
& RAW ensemble
& EMOS
& DM statistic
& \(p\)-value \\
\midrule
CRPS
& \(0.976\)
& \(\mathbf{0.850}\)
& \(\mathbf{15.853}\)
& \(\mathbf{1.43{\times}10^{-40}}\) \\

ES
& \(15.891\)
& \(\mathbf{15.223}\)
& \(\mathbf{6.953}\)
& \(\mathbf{2.63{\times}10^{-11}}\) \\

VS
& \(\mathbf{61607.16}\)
& \(69105.34\)
& \(-20.197\)
& \(4.10{\times}10^{-56}\) \\
\bottomrule
\end{tabular}
\caption{Mean CRPS, ES, and VS values for the raw ensemble and EMOS-based wind-speed forecasts over the validation period. CRPS values are evaluated from the univariate EMOS predictive distributions, while ES and VS values are based on EMOS-Q ensembles comprising 50 equidistant marginal quantiles. Two-sided DM tests use daily mean scores, with positive statistics favouring EMOS-based forecasts. Bold values indicate significant improvement at the \(5\%\) level.}
\label{crps_ws}
\end{table}

Ensemble copula coupling \citep[ECC;][]{SchefzikEtAl2013} is a two-step post-processing method that combines calibrated univariate predictive distributions with the dependence structure of the raw ensemble. Let $K$ denote the number of ensemble members and $D$ the number of forecast dimensions, such as stations. In the first step, $K$ values are obtained separately from each calibrated marginal predictive distribution. In the second step, these values are reordered according to the rank structure of the raw ensemble. The basic ECC variants differ only in how the marginal samples are generated. ECC-Q uses $K$ fixed equidistant quantiles, that is
\[
q_k=\frac{k}{K+1}, \qquad k=1,\ldots,K,
\]
and the corresponding marginal samples are given by
\[
\tilde f _k^{(d)}=F_d^{-1}(q_k), \qquad d=1,\ldots,D,
\]
where \(F_d^{-1}\) denotes the quantile function of the post-processed marginal predictive CDF \(F_d\) at dimension \(d\). In the case studies of Section \ref{sec5}, we also utilize these equidistant quantiles as a reference model that assumes the independence of the marginals, and is referred to as EMOS-Q. ECC-R uses $K$ independent random draws from each calibrated predictive distribution. The dual ensemble copula coupling (dECC) method of \cite{BenBouallegueEtAl2016} extends ECC-Q by correcting the raw ensemble dependence structure using historical forecast errors. First, an ECC-Q ensemble is generated from the calibrated marginal distributions. For each raw ensemble member, its deviation from the ensemble mean is transformed using the estimated forecast-error correlation matrix. The resulting correction is added to the corresponding ECC-Q member, producing an adjusted ensemble with a modified dependence structure. Finally, the calibrated marginal quantiles are rearranged according to the rank structure of this adjusted ensemble.

\begin{figure}[t]
    \centering
    \includegraphics[width=1\linewidth]{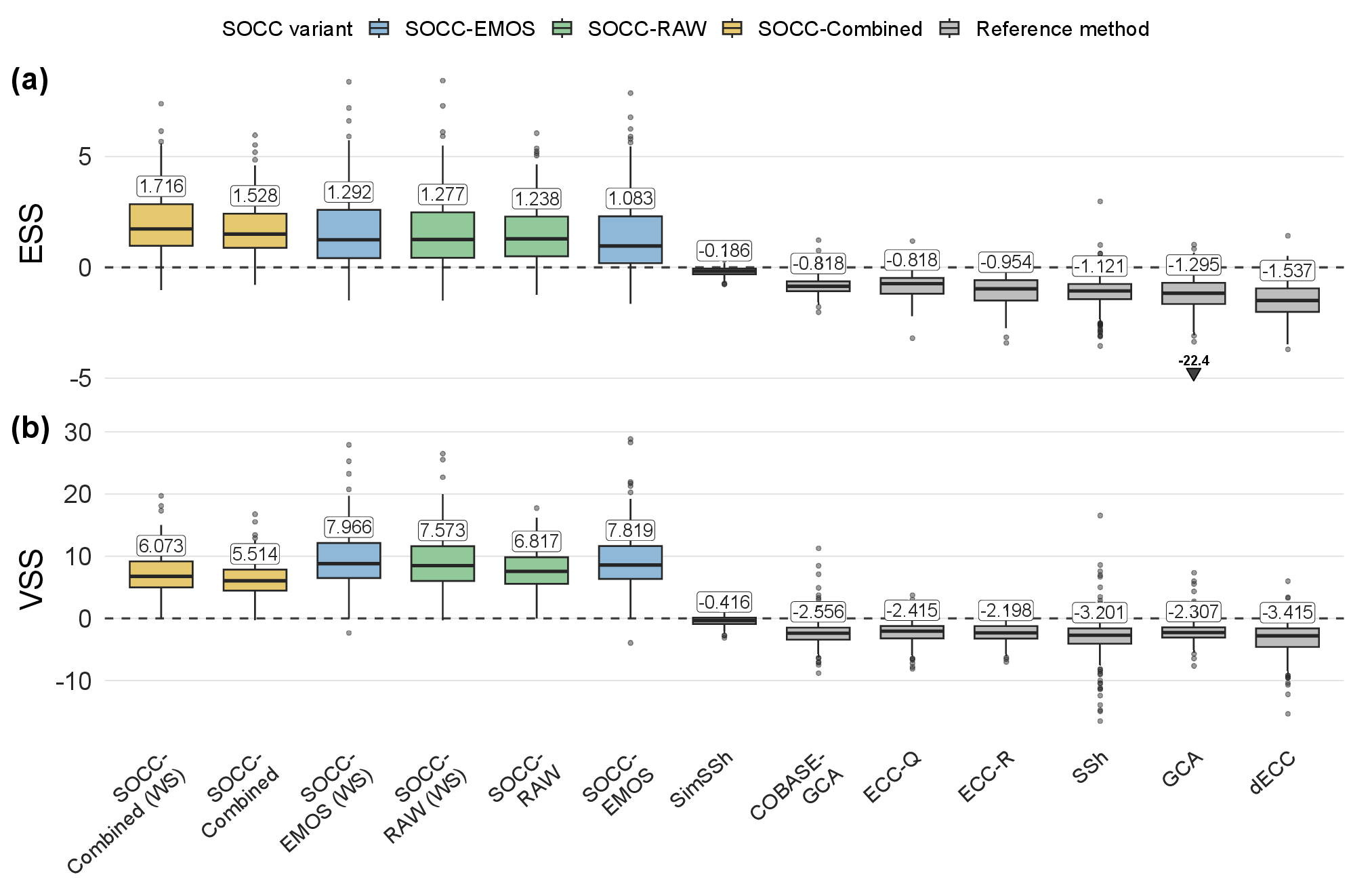}
\caption{Distributions of daily energy skill scores (ESS; panel a) and variogram skill scores (VSS; panel b) for wind-speed forecasts relative to mdSSh. Positive values indicate improvement, while the labels report the overall skill scores calculated from the corresponding mean ES and VS values. WS denotes the warm-start SOCC variants. For visual clarity, the underperforming EMOS-Q benchmark is omitted (ESS: \(-9.34\%\); VSS: \(-27.28\%\)).}
    \label{fig:ess_vss_wind}
\end{figure}

The Schaake shuffle \citep[SSh;][]{ClarkEtAl2004} uses historical observations to define the dependence template. For each forecast case, $K$ dates are randomly selected from a historical archive, and samples from the calibrated marginal predictive distributions are rearranged according to the rank structure of the corresponding observed fields. The similarity-based Schaake shuffle \citep[SimSSh;][]{Schefzik2016} selects historical dates with forecast situations similar to the current case instead of sampling them randomly. Similarity is evaluated using characteristics of the raw ensemble, such as its mean and spread. This allows the dependence template to reflect the expected meteorological situation more closely. The minimum-divergence Schaake shuffle \citep[mdSSh;][]{ScheuererEtAl2017} selects historical observations that are most consistent with the current marginal predictive distributions. Candidate dates are evaluated according to how well their observed values agree with prediction intervals derived from the calibrated marginals. The selected fields then provide the rank template used to rearrange the post-processed samples.

The Gaussian copula approach \citep[GCA;][]{MollerEtAl2013} estimates the dependence structure from historical observations and fits a multivariate distribution to the data by connecting marginal distributions with a Gaussian copula. In our case, the generated samples represent spatially connected ensemble forecasts. COBASE-GCA, a variant of the COBASE method of \cite{FlosEtAl2026}, uses the same Gaussian copula dependence structure but rearranges $K$ (in the present paper, fixed equidistant) marginal quantiles according to the ranks of the generated copula sample. Thus, GCA uses random marginal samples, whereas, in our case, to provide a fair comparison between competing methods, COBASE-GCA uses the same equidistant quantiles as ECC-Q and all the Schaake shuffle variants. Note that ECC-R, however, utilizes random draws from the calibrated CDFs.

\section{Results}
\label{sec5}

\begin{figure}[t]
    \centering
    \includegraphics[width=1\linewidth]{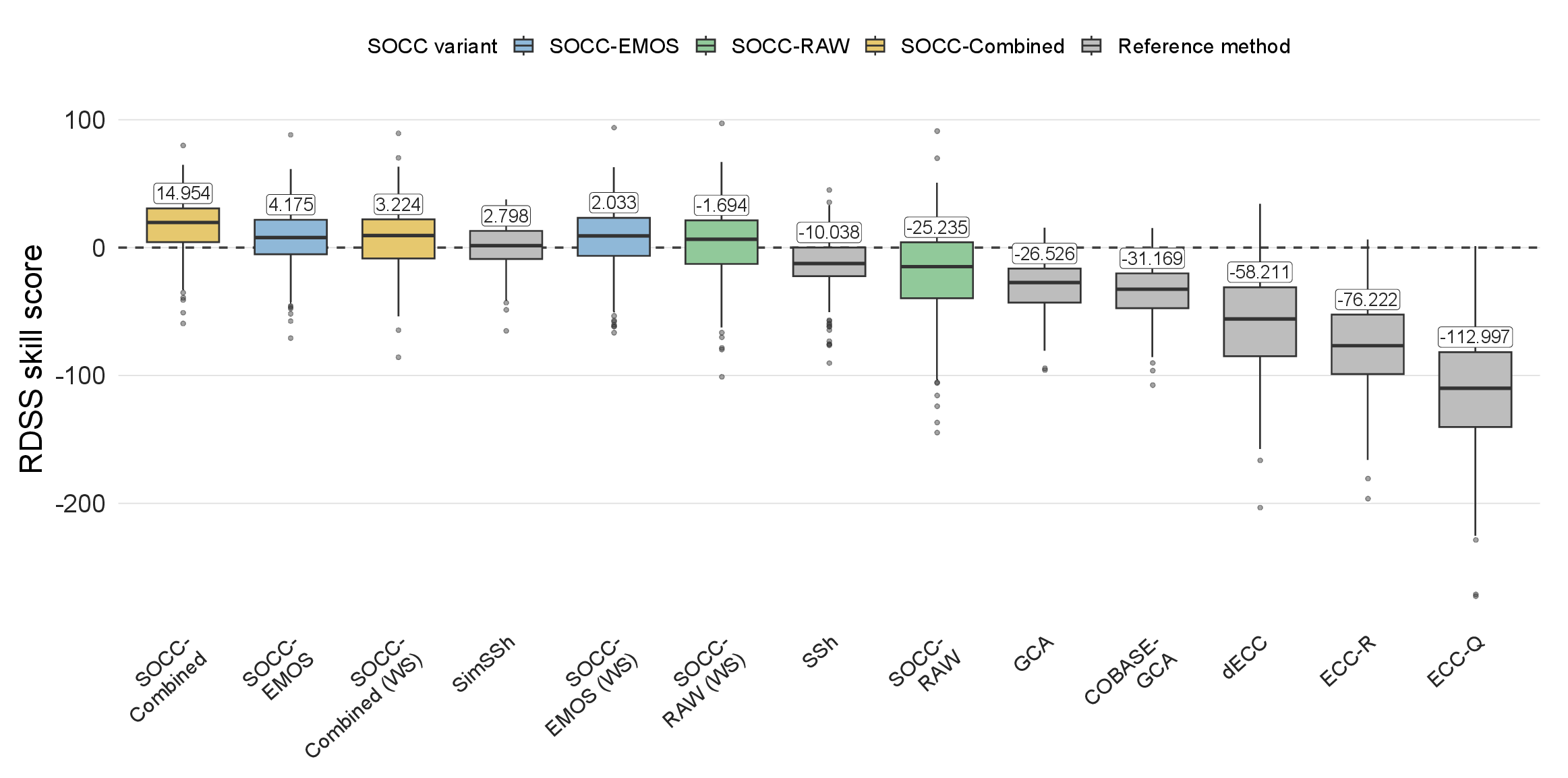}
\caption{Distributions of daily regularized Dawid--Sebastiani skill scores for wind speed forecasts relative to mdSSh. Positive values indicate improvement, while the labels report the overall skill scores calculated from the corresponding mean RDSS values. WS denotes the warm-start SOCC variants. For visual clarity, the underperforming EMOS-Q benchmark is omitted (overall RDSS skill: \(-500.590\%\)).}
    \label{fig:rdss_wind}
\end{figure}

This section presents the verification results for the spatial post-processing of wind speed and 2-m temperature, followed by a joint case study in which spatial and inter-variable dependencies are modelled simultaneously at common locations. For the first two case studies, we evaluate univariate predictive performance before assessing multivariate forecast quality and calibration. The third case study focuses exclusively on multivariate performance.

\subsection{Spatial dependence in wind speed forecasts}
\label{sec5.1}

Table \ref{crps_ws} presents the mean CRPS values of the raw ensemble forecasts and the post-processed EMOS predictive distributions for wind speed, together with the mean ES and VS values obtained when the EMOS distributions are represented by EMOS-Q ensembles. A two-sided DM test is used to assess the statistical significance of the difference in CRPS. As expected, univariate calibration reduces the mean CRPS by approximately \(13\%\), and the improvement is statistically significant. EMOS-Q also yields a lower mean ES than the raw ensemble. However, its higher mean VS suggests that univariate calibration alone does not adequately preserve the spatial dependence structure, motivating the application of the multivariate post-processing methods described in Sections \ref{sec4.2} and \ref{sec4.4}.

\begin{figure}
    \centering
    \includegraphics[width=1\linewidth]{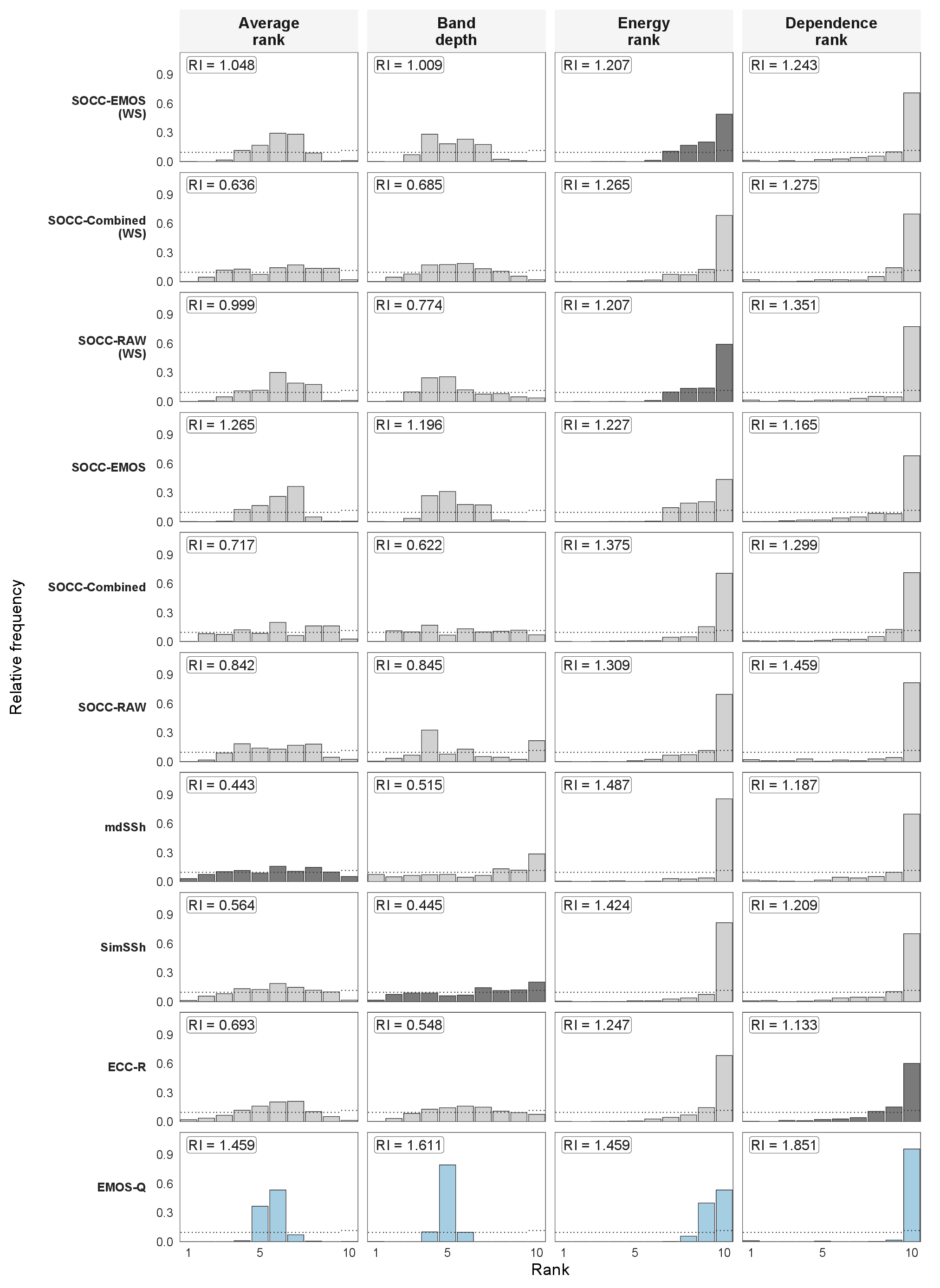}
    \caption{Multivariate rank histograms and corresponding reliability indices (RI) based on average rank, band depth rank, energy rank and dependence rank for post-processed wind-speed forecasts. Dark gray histograms indicate the methods with the lowest RI in each rank category. Dotted lines indicate uniform distributions.}
    \label{fig:ws_ranks}
\end{figure}

Figure \ref{fig:ess_vss_wind} shows boxplots of the distributions of daily ESS (panel a) and VSS (panel b) values for the competing multivariate forecasts relative to mdSSh. The overall ESS and VSS values are also reported. These are calculated using the skill-score formulation provided in Section \ref{sec3}, that is, they quantify the improvements in the overall mean ES and VS relative to the corresponding means of mdSSh. For a clearer comparison among the multivariate post-processing methods, EMOS-Q, which assumes independence among the marginals, is omitted. Based on Figure \ref{fig:ess_vss_wind}a, none of the benchmark methods, illustrated by gray boxplots, achieve a positive gain over mdSSh. The closest-performing forecast is SimSSh, followed by COBASE-GCA, ECC-Q, ECC-R, SSh, GCA, and finally dECC. In contrast, all SOCC variants outperform mdSSh, with overall gains ranging from \(1.083\%\) to \(1.716\%\). The strongest SOCC variant is warm-start SOCC-Combined, closely followed by its static variant. Warm-start SOCC-EMOS and SOCC-RAW also improve upon their static counterparts, with the difference being most pronounced for SOCC-EMOS. Figure \ref{fig:ess_vss_wind}b, which presents improvements in terms of the VS, shows the same general pattern, although the gains achieved by the SOCC variants are larger. Warm-start SOCC-EMOS achieves the largest gain of \(7.966\%\), closely followed by its static variant with a skill score of \(7.819\%\). These are followed by warm-start SOCC-RAW, SOCC-RAW, warm-start SOCC-Combined, and SOCC-Combined, with positive gains of \(7.573\%\), \(6.817\%\), \(6.073\%\), and \(5.514\%\), respectively. These improvements are statistically significant according to the corresponding DM tests reported in Table \ref{tab:dm_ws} in the Appendix.

\begin{table}[t]
\centering
\begin{tabular}{lrrrr}
\toprule
Score
& RAW ensemble
& EMOS
& DM statistic
& \(p\)-value \\
\midrule
CRPS
& \(1.129\)
& \(\mathbf{0.996}\)
& \(\mathbf{18.432}\)
& \(\mathbf{7.62{\times}10^{-50}}\) \\

ES
& \(11.844\)
& \(\mathbf{11.469}\)
& \(\mathbf{6.164}\)
& \(\mathbf{2.54{\times}10^{-9}}\) \\

VS
& \(\mathbf{17568.05}\)
& \(19310.17\)
& \(-25.101\)
& \(6.95{\times}10^{-73}\) \\
\bottomrule
\end{tabular}
\caption{Mean CRPS, ES, and VS values for the raw ensemble and EMOS-based temperature forecasts over the validation period. CRPS values are evaluated from the univariate EMOS predictive distributions, while ES and VS values are based on EMOS-Q ensembles comprising 50 equidistant marginal quantiles. Two-sided DM tests use daily mean scores, with positive statistics favouring EMOS-based forecasts. Bold values indicate significant improvement at the \(5\%\) level.}
\label{crps_t2m}
\end{table}

\begin{figure}[t]
    \centering
    \includegraphics[width=1\linewidth]{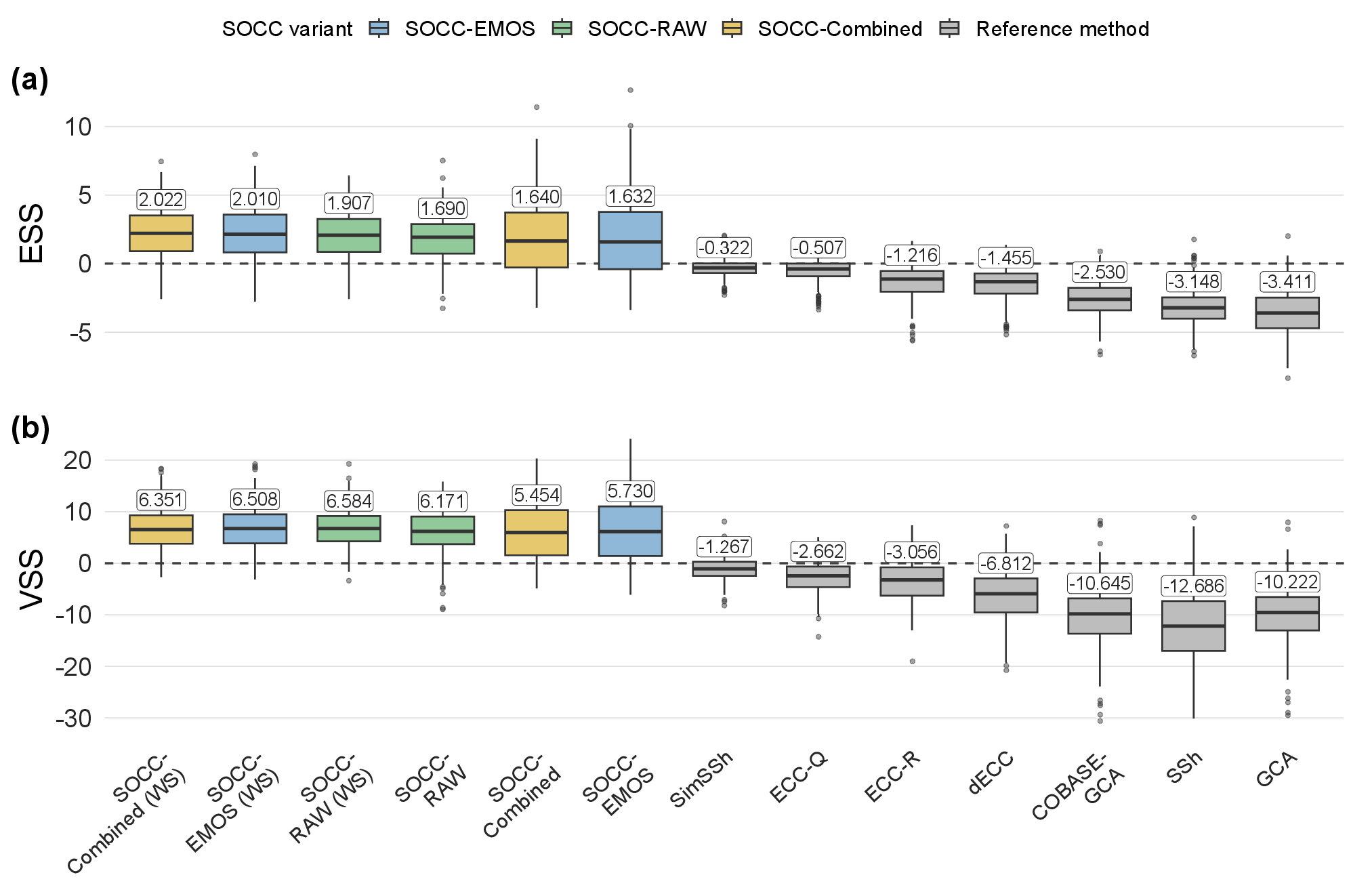}
\caption{Distributions of daily energy skill scores (ESS; panel a) and variogram skill scores (VSS; panel b) for temperature forecasts relative to mdSSh. Positive values indicate improvement, while the labels report the overall skill scores calculated from the corresponding mean ES and VS values. WS denotes the warm-start SOCC variants. For visual clarity, the underperforming EMOS-Q benchmark is omitted (mean ESS: \(-8.926\%\); mean VSS: \(-28.643\%\)).}
    \label{fig:ess_vss_t2m}
\end{figure}

Figure \ref{fig:rdss_wind} shows boxplots of the daily RDSS skill scores, with labels reporting the overall skill scores calculated from the corresponding mean RDSS values. Static SOCC-Combined achieves the largest improvement over mdSSh, followed by static SOCC-EMOS, warm-start SOCC-Combined, SimSSh, and warm-start SOCC-EMOS. All remaining forecasts perform worse than mdSSh in terms of overall RDSS. The warm-start updates reduce the overall RDSS skill of both SOCC-Combined and SOCC-EMOS. In contrast, warm-start updating substantially improves SOCC-RAW, although its overall skill score remains negative.

\begin{figure}[t]
    \centering
    \includegraphics[width=1\linewidth]{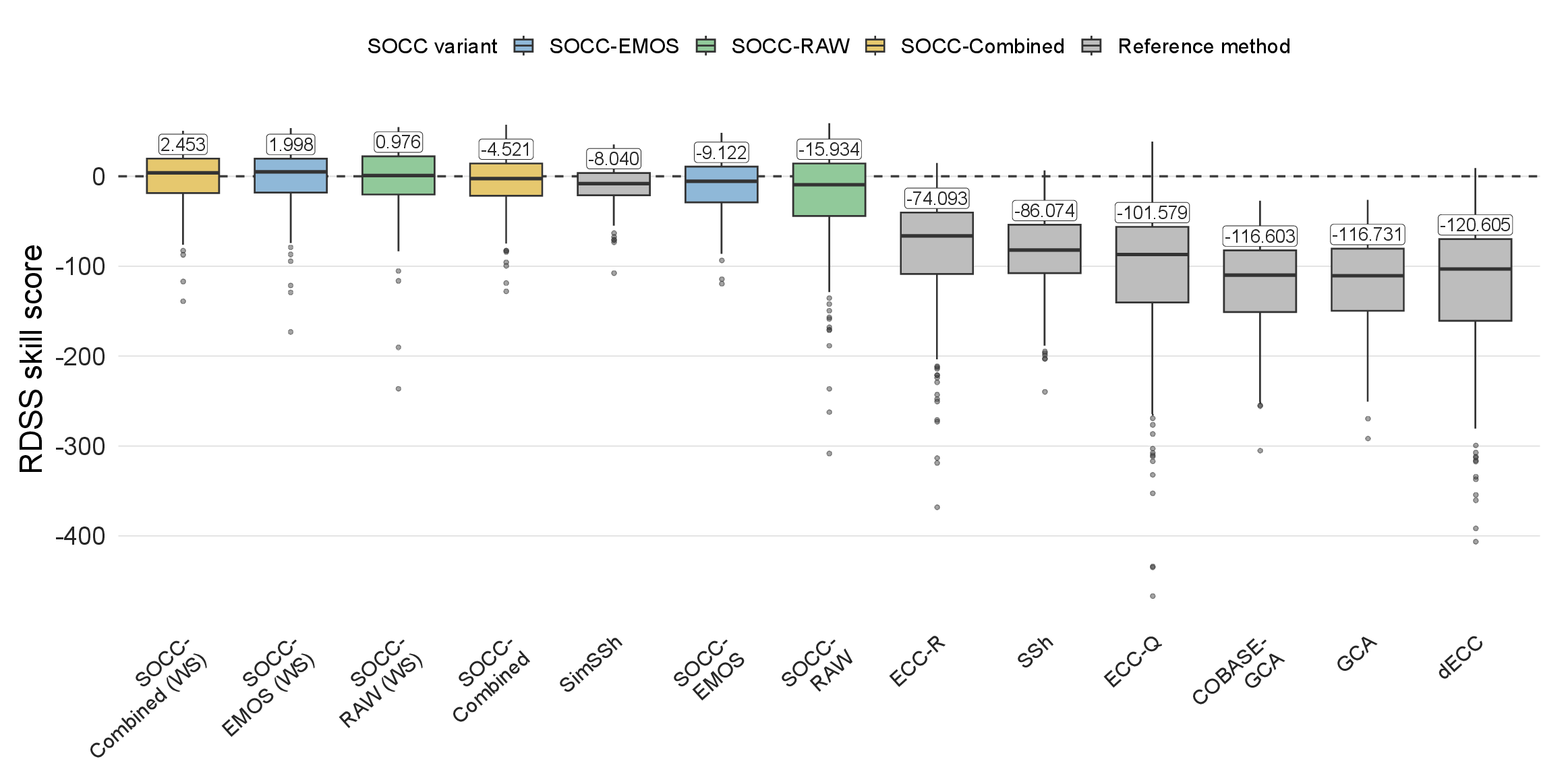}
\caption{Distributions of daily regularized Dawid--Sebastiani skill scores for 2-m temperature forecasts relative to mdSSh. Positive values indicate improvement, while the labels report the overall skill scores calculated from the corresponding mean RDSS values. WS denotes the warm-start SOCC variants. For visual clarity, the underperforming EMOS-Q benchmark is omitted (overall RDSS skill: \(-796.882\%\)).}
    \label{fig:rdss_t2m}
\end{figure}

Finally, Figure \ref{fig:ws_ranks} presents rank histograms for the SOCC variants and for the benchmark methods achieving the lowest RI in at least one of the investigated rank categories. EMOS-Q, which does not apply explicit multivariate dependence reordering, is also included as a benchmark. Except for the energy-rank histogram of mdSSh, all other displayed forecasts have lower RIs than EMOS-Q. However, identifying a single best-performing method is not straightforward. According to the average-rank and band-depth diagnostics, mdSSh and SimSSh produce the most uniform histograms, respectively. ECC-R has the lowest dependence-rank RI, while warm-start SOCC-EMOS and warm-start SOCC-RAW are tied for the lowest energy-rank RI. SOCC-Combined, whose loss function includes average-rank and band-depth penalties, produces substantially lower RIs for these two diagnostics than the other static SOCC variants. For the average-rank histogram, SOCC-Combined reduces the RI by approximately \(43\%\) relative to SOCC-EMOS and by \(15\%\) relative to SOCC-RAW. Among the warm-start variants, the corresponding reductions are approximately \(39\%\) and \(36\%\). However, these improvements come at a cost. The Combined variants have higher energy-rank RIs than the corresponding SOCC-EMOS and SOCC-RAW variants and rank second within their respective static and warm-start groups in terms of dependence-rank RI. Overall, these results demonstrate that improving the average-rank and band-depth histograms does not necessarily improve the energy- and dependence-rank histograms simultaneously.

\begin{figure}[p]
    \centering
    \includegraphics[
        width=\linewidth,
        height=0.88\textheight,
        keepaspectratio
    ]{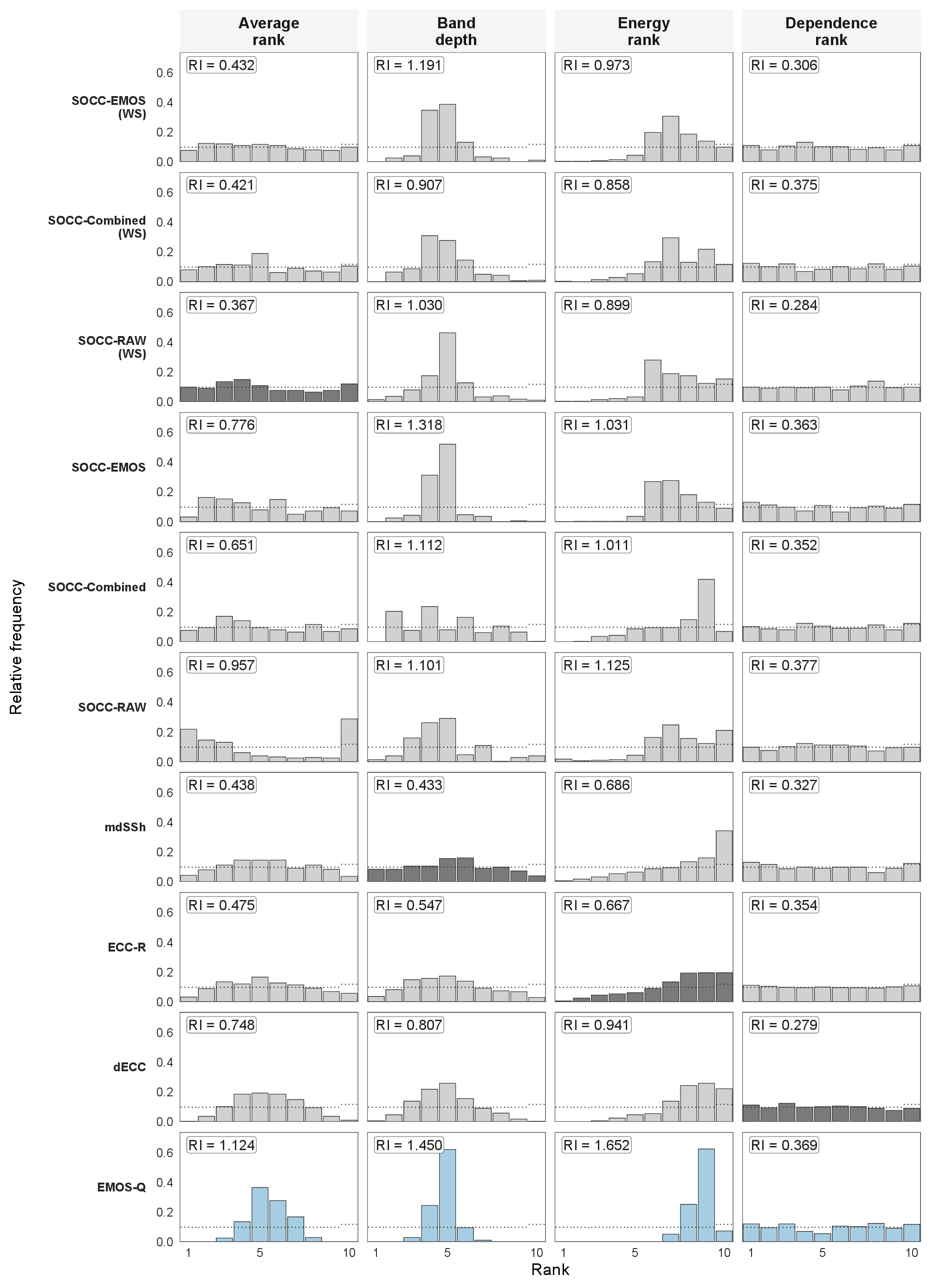}
    \caption{Multivariate rank histograms and corresponding reliability
    indices (RI) based on average rank, band depth rank, energy rank, and
    dependence rank for post-processed temperature forecasts. Dark gray
    histograms indicate the methods with the lowest RI in each rank category.
    Dotted lines indicate uniform distributions.}
    \label{fig:t2m_ranks}
\end{figure}

\subsection{Spatial dependence in temperature forecasts}
\label{sec5.2}

Table \ref{crps_t2m} presents the mean CRPS values of the raw and post-processed temperature forecasts. As in Table \ref{crps_ws}, the EMOS predictive distributions are represented by EMOS-Q ensembles when evaluating multivariate performance. Again, post-processing yields a lower mean CRPS and a lower mean ES, with corresponding skill scores of approximately \(11.8\%\) and \(3.2\%\), respectively. Both improvements are statistically significant. However, the higher mean VS of EMOS-Q suggests that marginal calibration alone does not adequately preserve the spatial dependence structure of the raw ensemble.

Figure \ref{fig:ess_vss_t2m} shows boxplots of daily ESS and VSS values relative to mdSSh. The overall ESS and VSS values are also reported, as in Figure \ref{fig:ess_vss_wind}. All multivariate forecasts outperform EMOS-Q, which is omitted from the figure to facilitate a clearer comparison among the competing methods. Figure \ref{fig:ess_vss_t2m}a shows that all SOCC variants outperform mdSSh, with overall ESS values ranging from \(1.632\%\) to \(2.022\%\), whereas none of the benchmark methods achieve positive skill. Figure \ref{fig:ess_vss_t2m}b shows a similar overall pattern. Here, warm-start SOCC-RAW achieves the largest positive gain over mdSSh, closely followed by the other warm-start variants. Overall, the warm-start SOCC variants outperform their static counterparts in terms of both ES and VS and generally exhibit narrower interquartile ranges, particularly in terms of VSS, suggesting more consistent daily performance. Among the benchmark methods, SimSSh comes closest to mdSSh. The largest losses are obtained by GCA and SSh in terms of ES, and by COBASE-GCA and SSh in terms of VS. The improvements of all SOCC variants over mdSSh are statistically significant, as confirmed by the two-sided DM test statistics and corresponding \(p\)-values reported in Table \ref{tab:dm_t2m} in the Appendix.

Figure \ref{fig:rdss_t2m} shows boxplots of the daily RDSS skill scores, with labels reporting the overall skill scores calculated from the corresponding mean RDSS values. Warm-start SOCC-Combined achieves the largest improvement over mdSSh, followed by warm-start SOCC-EMOS and warm-start SOCC-RAW. Static SOCC-Combined is the only static SOCC variant that outperforms SimSSh, although its overall skill score remains negative. However, the DM tests reported in Table \ref{tab:dm_t2m} indicate that none of the improvements achieved by the warm-start SOCC variants over mdSSh are statistically significant.

Finally, Figure \ref{fig:t2m_ranks} shows rank histograms based on the average-rank, band-depth-rank, energy-rank, and dependence-rank pre-rank functions for the competing forecasts. As in Figure \ref{fig:ws_ranks}, the figure includes the SOCC variants and the benchmark methods with the lowest RIs in at least one of the rank categories considered. EMOS-Q, which does not apply explicit multivariate dependence reconstruction, is also included for reference. All displayed multivariate forecasts have lower average-rank, band-depth-rank, and energy-rank RIs than EMOS-Q. The same holds for the dependence-rank RI, except for static SOCC-RAW and warm-start SOCC-Combined. Nevertheless, no single method performs best across all diagnostics. Warm-start SOCC-RAW has the most uniform average-rank histogram, while mdSSh, ECC-R, and dECC have the lowest RIs for band-depth rank, energy rank, and dependence rank, respectively.

Compared with static SOCC-EMOS, static SOCC-Combined has lower RIs across all four diagnostics, suggesting that adding the penalty terms to the loss function improves histogram uniformity. However, its band-depth RI remains slightly higher than that of static SOCC-RAW. The advantage of SOCC-Combined over SOCC-RAW is further reduced when warm-start optimization is used: warm-start SOCC-Combined has lower band-depth- and energy-rank RIs, but higher average-rank and dependence-rank RIs than warm-start SOCC-RAW.

\subsection{Joint spatial and inter-variable post-processing}
\label{sec5.3}

As mentioned earlier, we also investigate the simultaneous restoration of spatial and inter-variable dependence structures between temperature and wind speed. The analysis is based on the 24 stations for which forecasts and observations are available for both variables. We consider the benchmark methods introduced in the previous case studies as reference models. In the present setting, however, separate benchmark variants are constructed for the spatial and inter-variable dependence components, thereby providing strong component-specific reference forecasts against which the joint SOCC variants can be evaluated. The selected hyperparameters of the benchmark models for the spatial and inter-variable dependence components of this case study are reported in Table \ref{tab:selected_reference_hyperparameters} in the Appendix. The univariate performance of the wind speed and temperature forecasts has already been evaluated separately in Sections \ref{sec5.1} and \ref{sec5.2}. We therefore focus here exclusively on multivariate performance over the 24 co-located stations. Moreover, all methods considered in this case study preserve the calibrated marginal distributions.

Table \ref{tab:joint_component_skills} reports mean ESS,  mean VSS, and mean RDSS values for the spatial and inter-variable components separately. For the skill scores, as in the previous case studies, mdSSh serves as the reference method. For the RDSS, we report mean score values rather than skill scores because the ratio-based transformation can produce disproportionately large and difficult-to-interpret values when the reference score is close to zero, as is the case when evaluating inter-variable dependence. With respect to spatial dependence, warm-start SOCC-EMOS achieves the largest improvement over mdSSh in terms of both the ES and VS. Based on Table \ref{tab:dm_joint_components}, this improvement is also statistically significant. It should be noted, however, that mdSSh is not the strongest benchmark method in this setting. According to the mean RDSS, for which lower values indicate better predictive performance, warm-start SOCC-Combined performs best, which is also statistically significant along with the RDSS values of the static SOCC-Combined and the SOCC-EMOS variants. In the inter-variable setting, once again, the warm-start SOCC-EMOS achieves the largest gain over the reference in terms of ES, while SOCC-Combined is the most skilled based on the VS.  Similar to the spatial modelling RDSS results, warm-start SOCC-Combined has the lowest mean RDSS. All these gains are once again statistically significant based on Table \ref{tab:dm_joint_components}. These findings indicate a trade-off between the SOCC variants optimized solely in terms of the ES and VS and their combined counterparts. Introducing the multivariate-rank penalties consistently improves the RDSS relative to the corresponding SOCC-EMOS variants, indicating a better representation of the covariance structure. By contrast, warm-start SOCC-EMOS achieves the largest spatial ESS and VSS and the largest inter-variable ESS, whereas static SOCC-Combined achieves the largest inter-variable VSS.

\begin{table}[t]
\centering
\scriptsize
\setlength{\tabcolsep}{4pt}
\begin{tabular}{lrrrrrr}
\toprule
& \multicolumn{3}{c}{Spatial} & \multicolumn{3}{c}{Inter-variable} \\
\cmidrule(lr){2-4}\cmidrule(lr){5-7}
Forecast & ESS & VSS & Mean RDSS & ESS & VSS & Mean RDSS \\
\midrule
SOCC-Combined (WS) & 2.588 & 0.269 & \textbf{15.077} & 0.807 & 0.709 & \textbf{0.419} \\
SOCC-Combined & 3.593 & 2.118 & 41.733 & 0.888 & \textbf{1.060} & 3.336 \\
SOCC-EMOS (WS) & \textbf{3.976} & \textbf{3.692} & 62.364 & \textbf{1.161} & 0.930 & 6.722 \\
SOCC-EMOS & 3.493 & 3.574 & 54.977 & 0.475 & 1.036 & 7.305 \\
SOCC-RAW (WS) & 2.802 & -2.178 & 80.637 & 0.172 & -0.046 & 4.771 \\
SOCC-RAW & 2.929 & 0.760 & 123.485 & -0.751 & -0.226 & 6.729 \\
\midrule
dECC & 0.183 & 0.354 & 136.937 & -0.200 & 0.106 & 1.173 \\
ECC-Q & 0.674 & -0.277 & 120.781 & -0.263 & 0.026 & 1.499 \\
SimSSh & -0.225 & -0.081 & 82.042 & 0.204 & -0.003 & 0.508 \\
SSh & -0.808 & -0.020 & 106.381 & 0.141 & 0.093 & 0.511 \\
COBASE-GCA & -1.056 & -0.514 & 140.179 & 0.109 & 0.049 & 0.669 \\
ECC-R & -0.069 & -3.068 & 96.091 & -1.728 & -1.482 & 1.363 \\
GCA & -2.083 & -1.637 & 140.007 & -1.352 & -1.358 & 0.636 \\
EMOS-Q & -7.837 & -9.546 & 1830.573 & -3.822 & -0.707 & 96.468 \\
mdSSh &-- & -- & 73.798 & -- & -- & 1.012 \\
\bottomrule
\end{tabular}
\caption{Mean ESS and mean VSS relative to mdSSh and mean RDSS for the spatial and
inter-variable dependence components. Positive skill scores and lower RDSS
values indicate better performance. Bold values denote the best result in
each column.}
\label{tab:joint_component_skills}
\end{table}

\section{Discussion and conclusions}
\label{sec6}

This study introduced Score-Optimized Copula Coupling (SOCC), an optimization-based multivariate post-processing method for restoring dependence structures that may be lost during marginal calibration. We investigated both static and warm-start variants. The static approach uses a single optimized rank template for all forecast cases, whereas the warm-start approach sequentially updates the template using recent forecast--observation pairs. We also examined two approaches to template estimation, in which candidate rank structures were evaluated by rearranging either raw ensemble members or EMOS quantiles. Two loss functions were considered: one combining the energy and variogram scores and an extended version additionally incorporating average-rank and band-depth calibration penalties.

In the two variable-specific case studies, all SOCC variants significantly outperformed the strongest reference method in terms of both the ES and VS, with particularly pronounced improvements in the VS. The multivariate rank diagnostics nevertheless showed that improvements in these scores do not necessarily imply better performance in every aspect of multivariate calibration. The inclusion of rank penalties partly mitigated these deficiencies, although no method performed best across all diagnostics.

The joint case study showed that a single optimized rank template can simultaneously represent spatial and inter-variable dependence between wind speed and temperature. Different SOCC variants performed best according to the individual verification measures. The warm-start SOCC-EMOS variant produced the largest spatial ESS and VSS and the largest inter-variable ESS, whereas the static SOCC-Combined variant achieved the largest inter-variable VSS. Warm-start SOCC-Combined yielded the lowest mean RDSS for both dependence components.

The results based on the RDSS provide a more nuanced comparison of the loss functions. Optimization based only on the ES and VS generally produced larger improvements in these scores, whereas the combined loss generally yielded better performance in terms of the covariance-sensitive RDSS. This trade-off indicates that the loss function should be selected according to the aspects of multivariate forecast quality that are most important for the intended application. More generally, the results demonstrate that improving particular proper scores does not guarantee simultaneous improvement in all multivariate calibration diagnostics.

SOCC has both advantages and limitations compared with existing multivariate post-processing methods. Its optimization may become computationally demanding in high-dimensional settings. However, methods involving the estimation and manipulation of large covariance matrices, such as GCA, also scale unfavorably with dimension. A comparison of their theoretical computational complexities is provided in the Appendix.

In the case studies considered here, SOCC optimization remained computationally feasible without GPU acceleration. All static variants required less than six minutes, while the selected SOCC-EMOS and SOCC-Combined variants ran in under two minutes. Individual warm-start updates required only \(0.01\)--\(0.08\) seconds per forecast date. SOCC also has relatively few hyperparameters to tune, and its dependence reconstruction is nonparametric.

The current implementation uses a greedy optimization strategy and may therefore become trapped in a local minimum. Furthermore, the ES and VS assess only certain aspects of forecast quality, and emphasizing these scores may adversely affect rank-based calibration diagnostics or the representation of large-scale dependence structures. A static template also cannot fully represent temporal changes in dependence, although the warm-start updates allow it to adapt to more recent forecast--observation relationships.

Future research could improve the optimization procedure by identifying more informative initial templates and developing guided or adaptive rank-swapping strategies. Prioritizing swaps with a greater potential to reduce the loss could shorten the optimization while potentially improving forecast performance. Preliminary experiments conducted as part of this study also explored the incorporation of phase-space diagnostics into the loss function, including the alignment between the ensemble subspace and the realized forecast errors and the calibration of error amplitudes within this subspace \citep{WeiToth2003,LeutbecherLang2014}. Although these experiments demonstrated the feasibility of this extension, its systematic development and evaluation remain topics for future research. Further extensions could investigate global or stochastic optimization methods that are less susceptible to local minima, data-driven selection of loss weights, and dynamic templates that better capture temporal changes in dependence structures.

\section*{Funding statement}
The research was supported by the Hungarian National Research, Development and Innovation Office under Grant K142849 and by the EKÖP-26-4 University Research Scholarship Programme of the Ministry for Science and Technology, funded by the National Research,
Development and Innovation Fund.

\clearpage

\begin{appendices}

\setcounter{table}{0}
\renewcommand{\thetable}{A\arabic{table}}

\section{Hyperparameter selection for multivariate benchmark models}
\label{app:hyperparameters}

The hyperparameters of the benchmark multivariate post-processing methods
(described in Section \ref{sec4.4}) were selected separately for each case
study and, in the joint case study, for spatial and inter-variable dependence,
by minimizing the corresponding equally weighted loss function over the
validation period. The same selection procedure was applied to all competing
benchmark methods.

For the Schaake shuffle methods, the candidate values were \(1\), \(3\),
\(5\), and \(9\) years for the archive length, which refers to the length of
the historical period from which observations were obtained to construct the
dependence templates. We also considered \(30\), \(60\), \(90\), and \(183\)
days for the seasonal radius, restricting the potential candidates to dates
whose calendar-day distance from the forecast date does not exceed the
specified radius. In the case of SimSSh, which selects dates from the
historical archive whose ensemble forecasts most closely resemble those for
the validation dates, we considered different weights for the ensemble mean
and standard deviation. The candidate weights assigned to the ensemble mean
were \(0.25\), \(0.50\), and \(0.75\), with the remaining weight assigned to
the standard deviation. For mdSSh, the investigated prediction-interval
coverages were \(0.80\), \(0.90\), \(0.95\), and \(0.99\). The
dependence-estimation window for dECC, GCA, and COBASE-GCA was also selected
separately using the validation period. Here, dependence estimation refers to
the estimation of the correlation structure required by each method from
forecast errors or observations over a rolling window preceding the forecast
date. Candidate window lengths of \(30\), \(60\), \(90\), \(120\), and
\(180\) days were compared. Table
\ref{tab:selected_reference_hyperparameters} reports the selected values for
the wind-speed, temperature, and joint case studies.

\begin{table}[h]
\centering
\scriptsize
\setlength{\tabcolsep}{4pt}
\renewcommand{\arraystretch}{0.95}

\begin{adjustbox}{max width=\linewidth}
\begin{tabular}{llccc}
\toprule
Method & Parameter
& Wind speed
& Temperature
& Joint: spatial/inter-variable \\
\midrule

SSh
& Archive / radius (years/days)
& \(1/30\) & \(1/30\) & \(1/30\,;\,5/30\) \\

SimSSh
& Archive / radius; mean weight
& \(5/183;\ 0.75\)
& \(9/60;\ 0.75\)
& \(1/30;\ 0.50\, /\, 5/30;\ 0.25\) \\

mdSSh
& Archive / radius; coverage
& \(9/183;\ 0.80\)
& \(9/60;\ 0.99\)
& \(1/30;\ 0.99\, /\, 5/30;\ 0.80\) \\

\midrule

dECC
& Window (days)
& 30 & 120 & \(90/90\) \\

GCA
& Window (days)
& 60 & 30 & \(90/90\) \\

COBASE-GCA
& Window (days)
& 60 & 30 & \(90/90\) \\

\bottomrule
\end{tabular}
\end{adjustbox}

\caption{Selected hyperparameters for the reference methods. Archive/radius
entries report the archive length in years followed by the seasonal radius
in days. In the joint case study, paired entries are reported in the order
spatial/inter-variable. ECC-Q and ECC-R have no tuned hyperparameters and
are therefore omitted.}
\label{tab:selected_reference_hyperparameters}
\end{table}

\section{Results of the Diebold--Mariano test}

Tables \ref{tab:dm_ws}, \ref{tab:dm_t2m}, and \ref{tab:dm_joint_components} summarize the results of two-sided Diebold--Mariano tests comparing the investigated methods with mdSSh in terms of the ES, VS, and RDSS. The tests are based on paired daily score values, with the score differential defined such that positive DM statistics favour the competing method, whereas negative values favour mdSSh. Thus, a positive DM statistic with a two-sided \(p\)-value below \(0.05\) indicates a statistically significant improvement over mdSSh.

\setcounter{table}{0}
\renewcommand{\thetable}{B\arabic{table}}

\begin{table}[p]
\centering
\fontsize{6.5}{7.1}\selectfont
\setlength{\tabcolsep}{2.5pt}
\renewcommand{\arraystretch}{0.82}

\begin{tabular}{lcccccc}
\toprule
& \multicolumn{2}{c}{ES}
& \multicolumn{2}{c}{VS}
& \multicolumn{2}{c}{RDSS} \\
\cmidrule(lr){2-3}
\cmidrule(lr){4-5}
\cmidrule(lr){6-7}
Forecast
& DM statistic & \(p\)-value
& DM statistic & \(p\)-value
& DM statistic & \(p\)-value \\
\midrule

SOCC-EMOS (WS)
& \(\mathbf{16.756}\)
& \(\mathbf{7.95{\times}10^{-44}}\)
& \(\mathbf{54.657}\)
& \(\mathbf{4.76{\times}10^{-149}}\)
& \(1.235\)
& \(2.18{\times}10^{-1}\) \\

SOCC-Combined (WS)
& \(\mathbf{28.572}\)
& \(\mathbf{5.10{\times}10^{-84}}\)
& \(\mathbf{56.058}\)
& \(\mathbf{8.36{\times}10^{-152}}\)
& \(\mathbf{2.033}\)
& \(\mathbf{4.30{\times}10^{-2}}\) \\

SOCC-Combined
& \(\mathbf{28.941}\)
& \(\mathbf{3.66{\times}10^{-85}}\)
& \(\mathbf{50.522}\)
& \(\mathbf{1.47{\times}10^{-140}}\)
& \(\mathbf{11.584}\)
& \(\mathbf{1.71{\times}10^{-25}}\) \\

SOCC-EMOS
& \(\mathbf{13.370}\)
& \(\mathbf{1.06{\times}10^{-31}}\)
& \(\mathbf{50.730}\)
& \(\mathbf{5.34{\times}10^{-141}}\)
& \(\mathbf{3.068}\)
& \(\mathbf{2.37{\times}10^{-3}}\) \\

SOCC-RAW (WS)
& \(\mathbf{17.272}\)
& \(\mathbf{1.11{\times}10^{-45}}\)
& \(\mathbf{53.702}\)
& \(\mathbf{3.89{\times}10^{-147}}\)
& \(-0.908\)
& \(3.65{\times}10^{-1}\) \\

SOCC-RAW
& \(\mathbf{18.820}\)
& \(\mathbf{3.13{\times}10^{-51}}\)
& \(\mathbf{49.634}\)
& \(\mathbf{1.16{\times}10^{-138}}\)
& \(-9.715\)
& \(2.41{\times}10^{-19}\) \\

SimSSh
& \(-13.979\)
& \(7.34{\times}10^{-34}\)
& \(-6.936\)
& \(2.91{\times}10^{-11}\)
& \(\mathbf{2.782}\)
& \(\mathbf{5.79{\times}10^{-3}}\) \\

ECC-Q
& \(-24.411\)
& \(1.35{\times}10^{-70}\)
& \(-17.346\)
& \(6.04{\times}10^{-46}\)
& \(-26.408\)
& \(4.93{\times}10^{-77}\) \\

COBASE-GCA
& \(-36.958\)
& \(2.96{\times}10^{-108}\)
& \(-19.114\)
& \(2.81{\times}10^{-52}\)
& \(-23.795\)
& \(1.97{\times}10^{-68}\) \\

ECC-R
& \(-25.042\)
& \(1.09{\times}10^{-72}\)
& \(-19.235\)
& \(1.05{\times}10^{-52}\)
& \(-26.359\)
& \(7.09{\times}10^{-77}\) \\

SSh
& \(-25.235\)
& \(2.51{\times}10^{-73}\)
& \(-15.556\)
& \(1.67{\times}10^{-39}\)
& \(-8.644\)
& \(4.75{\times}10^{-16}\) \\

GCA
& \(-6.039\)
& \(5.05{\times}10^{-9}\)
& \(-21.353\)
& \(3.65{\times}10^{-60}\)
& \(-22.281\)
& \(2.72{\times}10^{-63}\) \\

dECC
& \(-28.305\)
& \(3.48{\times}10^{-83}\)
& \(-17.379\)
& \(4.57{\times}10^{-46}\)
& \(-21.925\)
& \(4.58{\times}10^{-62}\) \\

EMOS-Q
& \(-103.444\)
& \(5.21{\times}10^{-221}\)
& \(-42.352\)
& \(5.20{\times}10^{-122}\)
& \(-40.725\)
& \(9.25{\times}10^{-118}\) \\

\bottomrule
\end{tabular}

\caption{Diebold--Mariano test results for wind-speed forecasts relative
to mdSSh. Bold values indicate significant improvement at the \(5\%\) level.}
\label{tab:dm_ws}

\vspace{0.4em}

\begin{tabular}{lcccccc}
\toprule
& \multicolumn{2}{c}{ES}
& \multicolumn{2}{c}{VS}
& \multicolumn{2}{c}{RDSS} \\
\cmidrule(lr){2-3}
\cmidrule(lr){4-5}
\cmidrule(lr){6-7}
Forecast
& DM statistic & \(p\)-value
& DM statistic & \(p\)-value
& DM statistic & \(p\)-value \\
\midrule

SOCC-EMOS (WS)
& \(\mathbf{18.696}\)
& \(\mathbf{8.71{\times}10^{-51}}\)
& \(\mathbf{25.890}\)
& \(\mathbf{1.80{\times}10^{-75}}\)
& \(1.079\)
& \(2.82{\times}10^{-1}\) \\

SOCC-Combined (WS)
& \(\mathbf{19.539}\)
& \(\mathbf{8.71{\times}10^{-54}}\)
& \(\mathbf{26.102}\)
& \(\mathbf{3.68{\times}10^{-76}}\)
& \(1.421\)
& \(1.56{\times}10^{-1}\) \\

SOCC-Combined
& \(\mathbf{11.139}\)
& \(\mathbf{5.27{\times}10^{-24}}\)
& \(\mathbf{17.279}\)
& \(\mathbf{1.05{\times}10^{-45}}\)
& \(-2.521\)
& \(1.23{\times}10^{-2}\) \\

SOCC-EMOS
& \(\mathbf{10.050}\)
& \(\mathbf{2.01{\times}10^{-20}}\)
& \(\mathbf{16.296}\)
& \(\mathbf{3.62{\times}10^{-42}}\)
& \(-5.168\)
& \(4.57{\times}10^{-7}\) \\

SOCC-RAW (WS)
& \(\mathbf{20.003}\)
& \(\mathbf{1.99{\times}10^{-55}}\)
& \(\mathbf{28.082}\)
& \(\mathbf{1.74{\times}10^{-82}}\)
& \(0.476\)
& \(6.34{\times}10^{-1}\) \\

SOCC-RAW
& \(\mathbf{17.320}\)
& \(\mathbf{7.44{\times}10^{-46}}\)
& \(\mathbf{21.631}\)
& \(\mathbf{3.94{\times}10^{-61}}\)
& \(-5.540\)
& \(7.12{\times}10^{-8}\) \\

SimSSh
& \(-8.839\)
& \(1.21{\times}10^{-16}\)
& \(-8.908\)
& \(7.52{\times}10^{-17}\)
& \(-6.996\)
& \(2.03{\times}10^{-11}\) \\

ECC-Q
& \(-10.874\)
& \(4.03{\times}10^{-23}\)
& \(-13.196\)
& \(4.32{\times}10^{-31}\)
& \(-19.341\)
& \(4.37{\times}10^{-53}\) \\

COBASE-GCA
& \(-33.118\)
& \(1.30{\times}10^{-97}\)
& \(-24.971\)
& \(1.86{\times}10^{-72}\)
& \(-29.771\)
& \(1.03{\times}10^{-87}\) \\

ECC-R
& \(-19.370\)
& \(3.46{\times}10^{-53}\)
& \(-13.855\)
& \(2.03{\times}10^{-33}\)
& \(-19.547\)
& \(8.12{\times}10^{-54}\) \\

SSh
& \(-37.756\)
& \(2.25{\times}10^{-110}\)
& \(-24.465\)
& \(8.92{\times}10^{-71}\)
& \(-24.587\)
& \(3.52{\times}10^{-71}\) \\

GCA
& \(-36.096\)
& \(6.26{\times}10^{-106}\)
& \(-25.585\)
& \(1.78{\times}10^{-74}\)
& \(-29.250\)
& \(4.08{\times}10^{-86}\) \\

dECC
& \(-20.490\)
& \(3.82{\times}10^{-57}\)
& \(-18.573\)
& \(2.39{\times}10^{-50}\)
& \(-19.947\)
& \(3.14{\times}10^{-55}\) \\

EMOS-Q
& \(-86.675\)
& \(1.17{\times}10^{-200}\)
& \(-33.964\)
& \(5.07{\times}10^{-100}\)
& \(-41.094\)
& \(6.41{\times}10^{-119}\) \\

\bottomrule
\end{tabular}

\caption{Diebold--Mariano test results for 2-m temperature forecasts
relative to mdSSh. Bold values indicate significant improvement at the
\(5\%\) level.}
\label{tab:dm_t2m}

\vspace{0.4em}

\begin{tabular}{lcccccc}
\toprule
& \multicolumn{2}{c}{ES} & \multicolumn{2}{c}{VS} & \multicolumn{2}{c}{RDSS} \\
\cmidrule(lr){2-3}\cmidrule(lr){4-5}\cmidrule(lr){6-7}
Forecast & DM statistic & \(p\)-value & DM statistic & \(p\)-value & DM statistic & \(p\)-value \\
\midrule
\multicolumn{7}{l}{\textit{(a) Spatial dependence}} \\
\addlinespace[2pt]
SOCC-Combined (WS) & \(\mathbf{42.194}\) & \(\mathbf{2.24{\times}10^{-121}}\) & \(1.126\) & \(2.61{\times}10^{-1}\) & \(\mathbf{20.568}\) & \(\mathbf{2.39{\times}10^{-57}}\) \\
SOCC-Combined & \(\mathbf{44.358}\) & \(\mathbf{1.55{\times}10^{-126}}\) & \(\mathbf{10.202}\) & \(\mathbf{6.66{\times}10^{-21}}\) & \(\mathbf{13.758}\) & \(\mathbf{4.75{\times}10^{-33}}\) \\
SOCC-EMOS (WS) & \(\mathbf{47.760}\) & \(\mathbf{2.82{\times}10^{-134}}\) & \(\mathbf{17.636}\) & \(\mathbf{6.13{\times}10^{-47}}\) & \(\mathbf{3.649}\) & \(\mathbf{3.15{\times}10^{-4}}\) \\
SOCC-EMOS & \(\mathbf{40.622}\) & \(\mathbf{1.67{\times}10^{-117}}\) & \(\mathbf{19.396}\) & \(\mathbf{3.24{\times}10^{-53}}\) & \(\mathbf{7.361}\) & \(\mathbf{2.17{\times}10^{-12}}\) \\
SOCC-RAW (WS) & \(\mathbf{37.155}\) & \(\mathbf{1.43{\times}10^{-108}}\) & \(-6.153\) & \(2.71{\times}10^{-9}\) & \(-2.110\) & \(3.58{\times}10^{-2}\) \\
SOCC-RAW & \(\mathbf{38.920}\) & \(\mathbf{3.48{\times}10^{-113}}\) & \(\mathbf{2.922}\) & \(\mathbf{3.77{\times}10^{-3}}\) & \(-9.062\) & \(2.60{\times}10^{-17}\) \\
dECC & \(\mathbf{3.107}\) & \(\mathbf{2.09{\times}10^{-3}}\) & \(1.967\) & \(5.02{\times}10^{-2}\) & \(-12.172\) & \(1.67{\times}10^{-27}\) \\
ECC-Q & \(\mathbf{12.086}\) & \(\mathbf{3.33{\times}10^{-27}}\) & \(-1.313\) & \(1.90{\times}10^{-1}\) & \(-9.227\) & \(8.09{\times}10^{-18}\) \\
SimSSh & \(-7.785\) & \(1.46{\times}10^{-13}\) & \(-1.246\) & \(2.14{\times}10^{-1}\) & \(-8.374\) & \(2.98{\times}10^{-15}\) \\
SSh & \(-16.006\) & \(4.36{\times}10^{-41}\) & \(-0.158\) & \(8.75{\times}10^{-1}\) & \(-15.569\) & \(1.63{\times}10^{-39}\) \\
COBASE-GCA & \(-18.950\) & \(1.24{\times}10^{-51}\) & \(-3.616\) & \(3.57{\times}10^{-4}\) & \(-22.968\) & \(1.23{\times}10^{-65}\) \\
ECC-R & \(-1.052\) & \(2.94{\times}10^{-1}\) & \(-11.092\) & \(7.85{\times}10^{-24}\) & \(-5.694\) & \(3.21{\times}10^{-8}\) \\
GCA & \(-29.057\) & \(2.22{\times}10^{-85}\) & \(-11.670\) & \(8.75{\times}10^{-26}\) & \(-22.464\) & \(6.46{\times}10^{-64}\) \\
EMOS-Q & \(-92.406\) & \(1.86{\times}10^{-207}\) & \(-15.719\) & \(4.69{\times}10^{-40}\) & \(-37.652\) & \(6.97{\times}10^{-110}\) \\
\midrule
\multicolumn{7}{l}{\textit{(b) Inter-variable dependence}} \\
\addlinespace[2pt]
SOCC-Combined (WS) & \(\mathbf{21.591}\) & \(\mathbf{6.52{\times}10^{-61}}\) & \(\mathbf{11.658}\) & \(\mathbf{9.67{\times}10^{-26}}\) & \(\mathbf{10.173}\) & \(\mathbf{8.24{\times}10^{-21}}\) \\
SOCC-Combined & \(\mathbf{16.279}\) & \(\mathbf{4.56{\times}10^{-42}}\) & \(\mathbf{11.457}\) & \(\mathbf{4.65{\times}10^{-25}}\) & \(-8.411\) & \(2.33{\times}10^{-15}\) \\
SOCC-EMOS (WS) & \(\mathbf{21.306}\) & \(\mathbf{6.36{\times}10^{-60}}\) & \(\mathbf{11.170}\) & \(\mathbf{4.29{\times}10^{-24}}\) & \(-10.016\) & \(2.65{\times}10^{-20}\) \\
SOCC-EMOS & \(\mathbf{6.885}\) & \(\mathbf{3.99{\times}10^{-11}}\) & \(\mathbf{9.412}\) & \(\mathbf{2.14{\times}10^{-18}}\) & \(-10.338\) & \(2.41{\times}10^{-21}\) \\
SOCC-RAW (WS) & \(\mathbf{2.538}\) & \(\mathbf{1.17{\times}10^{-2}}\) & \(-0.300\) & \(7.64{\times}10^{-1}\) & \(-10.373\) & \(1.85{\times}10^{-21}\) \\
SOCC-RAW & \(-8.113\) & \(1.71{\times}10^{-14}\) & \(-1.060\) & \(2.90{\times}10^{-1}\) & \(-10.546\) & \(5.05{\times}10^{-22}\) \\
dECC & \(-3.663\) & \(3.00{\times}10^{-4}\) & \(1.137\) & \(2.57{\times}10^{-1}\) & \(-2.066\) & \(3.98{\times}10^{-2}\) \\
ECC-Q & \(-4.866\) & \(1.93{\times}10^{-6}\) & \(0.311\) & \(7.56{\times}10^{-1}\) & \(-5.181\) & \(4.31{\times}10^{-7}\) \\
SimSSh & \(\mathbf{6.605}\) & \(\mathbf{2.09{\times}10^{-10}}\) & \(-0.057\) & \(9.55{\times}10^{-1}\) & \(\mathbf{10.375}\) & \(\mathbf{1.82{\times}10^{-21}}\) \\
SSh & \(\mathbf{4.151}\) & \(\mathbf{4.43{\times}10^{-5}}\) & \(1.075\) & \(2.83{\times}10^{-1}\) & \(\mathbf{10.361}\) & \(\mathbf{2.03{\times}10^{-21}}\) \\
COBASE-GCA & \(\mathbf{3.518}\) & \(\mathbf{5.09{\times}10^{-4}}\) & \(0.834\) & \(4.05{\times}10^{-1}\) & \(\mathbf{6.162}\) & \(\mathbf{2.59{\times}10^{-9}}\) \\
ECC-R & \(-28.211\) & \(9.34{\times}10^{-83}\) & \(-13.740\) & \(5.50{\times}10^{-33}\) & \(-4.120\) & \(5.03{\times}10^{-5}\) \\
GCA & \(-28.826\) & \(1.15{\times}10^{-84}\) & \(-13.595\) & \(1.80{\times}10^{-32}\) & \(\mathbf{6.681}\) & \(\mathbf{1.33{\times}10^{-10}}\) \\
EMOS-Q & \(-33.886\) & \(1.29{\times}10^{-99}\) & \(-2.854\) & \(4.65{\times}10^{-3}\) & \(-25.356\) & \(1.30{\times}10^{-73}\) \\
\bottomrule
\end{tabular}

\caption{Diebold--Mariano test results for the spatial and inter-variable
components of the joint forecasts relative to mdSSh. Bold values indicate
significant improvement at the \(5\%\) level.}
\label{tab:dm_joint_components}

\end{table}

\clearpage
\section{Computational cost, scalability and stability of SOCC variants}

\setcounter{table}{0}
\renewcommand{\thetable}{C\arabic{table}}

Besides forecast accuracy, computational efficiency is an important consideration when evaluating post-processing methods. ECC variants (such as ECC-Q and ECC-R) are computationally inexpensive, as they only require the sorting of calibrated marginal samples and the subsequent reordering based on the rank of \(K\) raw ensemble forecasts. In the case of \(D\) marginal distributions, the computational complexity of ECC variants is  \(O(DK\log K)\), where \(K\log K\) is the complexity of the most efficient sorting algorithm.

For SOCC, the initialization requires \(O(TDK^2+TD^2K)\) operations when calculating the ES and VS based on \(K\) ensemble members, \(D\) stations, and \(T\) training days. The asymptotic time complexity of each rank swap is \(O(T(D+K))\), as after a proposed rank swap, only \(2(K-2)T\) ensemble-member distances need to be recalculated for the ES, while only the \(T(D-1)\) station pairs involving the selected station need to be updated for the VS. As a result, for \(P\) proposals, its overall complexity is \(O\!\left(TDK^2+TD^2K+PT(D+K)\right).\)
Although static SOCC can be computationally demanding because it uses a large number of proposals, its asymptotic dependence on \(D\) is quadratic when \(T\), \(K\), and \(P\) are fixed.

In comparison, for the GCA, the estimation and decomposition of the covariance matrix and the generation of \(K\) correlated samples result in an overall complexity of \(O(TD^2+D^3+D^2K)\), where \(O(TD^2)\) corresponds to estimating the \(D\times D\) covariance matrix from \(T\) training days, \(O(D^3)\) to its matrix decomposition, and \(O(D^2K)\) to generating the \(K\) correlated samples. As a result, GCA and COBASE-GCA can be more computationally efficient in lower-dimensional settings, however, this advantage diminishes as the number of marginal dimensions increases. However, while the computational complexities of GCA and COBASE-GCA are linear in \(K\), that of SOCC contains both linear and quadratic terms.

Table \ref{tab:socc_computational_times} summarizes the runtimes of the SOCC models investigated in the case studies of Section \ref{sec5}. The computations were performed on a computer equipped with an Intel Core i7-11700KF processor (8 cores and 16 threads) and 32 GB of RAM. The models were implemented in \texttt{R}, with the computationally intensive SOCC optimization implemented in \texttt{C++}. Static template optimization and individual warm-start updates were executed on a single thread. The reported times are medians over 20 independent runs, while the total runtime of each warm-start variant is estimated as the sum of the corresponding median static optimization time and the median daily update time multiplied by the number of forecast cases in the evaluation period.

\begin{table}[h]
\centering
\scriptsize
\setlength{\tabcolsep}{3.5pt}
\renewcommand{\arraystretch}{0.9}
\begin{tabular}{lccccccccc}
\toprule
& \multicolumn{3}{c}{\shortstack{Wind speed\\(spatial)}}
& \multicolumn{3}{c}{\shortstack{Temperature\\(spatial)}}
& \multicolumn{3}{c}{\shortstack{Temperature and wind speed\\
(spatial and inter-variable)}} \\
\cmidrule(lr){2-4}
\cmidrule(lr){5-7}
\cmidrule(lr){8-10}
SOCC variant
& Static & Total & Update
& Static & Total & Update
& Static & Total & Update \\
\midrule
SOCC-EMOS
& 1.77 & 23.69 & 0.08
& 1.80 & 10.02 & 0.03
& 0.64 & 3.38 & 0.01 \\

SOCC-RAW
& 342.96 & 364.87 & 0.08
& 191.68 & 199.90 & 0.03
& 1.99 & 7.47 & 0.02 \\

SOCC-Combined
& 3.64 & 11.86 & 0.03
& 1.83 & 10.05 & 0.03
& 0.67 & 4.78 & 0.02 \\
\bottomrule
\end{tabular}

\caption{SOCC optimization times in seconds, reported as medians over
20 independent runs. ``Static'' denotes the initial template optimization,
``Total'' the estimated total runtime of the warm-start variant over the
evaluation period, including the static fit and all daily updates, and
``Update'' the runtime of one daily update.}
\label{tab:socc_computational_times}
\end{table}

To assess the practical benefit and stochastic stability of the static SOCC
optimization, we repeated the wind-speed SOCC-EMOS experiment using 20
different random seeds. For each seed, we also evaluated an unoptimized
template obtained by independently permuting the ranks \(1,\ldots,K\) in
each dimension. All other settings, including the EMOS marginal forecasts,
the 30-day training window, the verification period, and \(\alpha=0.5\),
were kept unchanged. Averaged over the 20 repetitions, the optimized
SOCC-EMOS templates achieved an ESS of \(1.11\%\) and a VSS of \(7.60\%\)
relative to the corresponding random templates. The optimized template
outperformed the random template for every seed according to both scores,
and all improvements were statistically significant based on two-sided
Diebold--Mariano tests at the \(5\%\) significance level (results not shown). These results indicate that the improvement is
systematic rather than being driven by a favourable realization of the
stochastic optimization. Moreover, the coefficients of variation of the seed-specific mean scores of SOCC-EMOS were only \(0.0225\%\) for the
ES and \(0.0479\%\) for the VS, indicating that the static SOCC optimization
is highly stable with respect to the random seed.
\end{appendices}

\bibliographystyle{plainnat}
\bibliography{references}

\end{document}